\documentclass[aps,prl,reprint,showpacs,superscriptaddress]{revtex4-1}
\usepackage[latin9]{inputenc}
\usepackage[a4paper]{geometry}
\usepackage{color}
\usepackage{amsmath}
\usepackage{amssymb}
\usepackage{graphicx}

\makeatletter
\usepackage{mathrsfs}
\usepackage{enumitem}
\usepackage{tabu}
\usepackage{standalone}
\usepackage{import}

\newcommand{\ii}{\text{i}}

\newcommand{\lyxdot}{.}

\allowdisplaybreaks[4]

\newcommand{\thickhline}{%
    \noalign {\ifnum 0=`}\fi \hrule height 1.25pt
    \futurelet \reserved@a \@xhline
}
\newcolumntype{"}{@{\hskip\tabcolsep\vrule width 1.25pt\hskip\tabcolsep}}

\usepackage{graphicx}
\usepackage{dcolumn}
\usepackage[tight]{subfigure}
\usepackage{amsmath}
\usepackage{verbatim}
\usepackage{color}
\usepackage{bm} 
\usepackage{bbm}
\usepackage{natbib}
\usepackage{xspace}
\usepackage{marginnote}
\usepackage{mathtools}
\usepackage{dsfont}
\usepackage{hyperref} \hypersetup{colorlinks=true,linktoc=all,linkcolor=blue,breaklinks=true,citecolor=blue,urlcolor=blue}
\usepackage{etoolbox}
\robustify{\uparrow}
\robustify{\downarrow}
\robustify{\sum}
\robustify{\int}
\robustify{\nonumber}
\robustify{\cite}
\robustify{\footnote}
\newrobustcmd{\Figure}[2]{
  \begin{figure}[ht]
    \includegraphics[width=1.0\linewidth]{#1}
    \caption{#2}
  \end{figure}
}
\expandafter\newrobustcmd\csname Figure2\endcsname[3]{
\begin{figure}[ht]
  \includegraphics[width=0.9\linewidth]{#1}
  \\
  \includegraphics[width=0.9\linewidth]{#2}
  \caption{#3}
\end{figure}
}
\renewrobustcmd{\Re}{{\text{Re}}}
\renewrobustcmd{\Im}{{\text{Im}}}
\newrobustcmd{\eff}{\text{eff}} 
\newrobustcmd{\dagtot}{{\dag_\tot}}
\newrobustcmd{\dagres}{{\dag_\res}}
\newrobustcmd{\Ttot}{{\text{T}_\tot}}
\newrobustcmd{\Tres}{{\text{T}_\res}}
\newrobustcmd{\T}{{\text{T}}}
\newrobustcmd{\tot}{\text{tot}}
\newrobustcmd{\tun}{\text{T}}
\newrobustcmd{\res}{\text{R}}
\newrobustcmd{\un}{\text{i}}   
\newrobustcmd{\In}{\text{0}}   
\newrobustcmd{\state}{inverted stationary state\xspace}   
\newrobustcmd{\K}{\mathcal{K}}
\newrobustcmd{\D}{\mathcal{I}}
\newrobustcmd{\bJ}{\bar{J}}
\newrobustcmd{\tJ}{\tilde{J}}
\newrobustcmd{\bG}{\bar{G}}
\newrobustcmd{\tG}{\tilde{G}}
\renewrobustcmd{\P}{\mathcal{P}}   
\newrobustcmd{\W}{\tilde{W}}
\newrobustcmd{\Temp}{T}
\newrobustcmd{\dual}[1]{\bar{#1}}
\newrobustcmd{\one}{\mathds{1}}
\newrobustcmd{\ket}[1]{|#1\rangle}
\newrobustcmd{\bra}[1]{\langle#1|}
\newrobustcmd{\brkt}[1]{\langle #1 \rangle}
\newrobustcmd{\braket}[2]{\langle #1 | #2 \rangle}
\newrobustcmd{\Ket}[1]{\bm{|}#1\bm{)}}
\newrobustcmd{\Bra}[1]{\bm{(}#1\bm{|}}
\newrobustcmd{\Braket}[2]{\bm{(}#1\bm{|}#2\bm{)}}
\newrobustcmd{\Brkt}[1]{\bm{(} #1 \bm{)}}
\newrobustcmd{\op}[1]{\hat{#1}}
\DeclareMathOperator{\Tr}{Tr}
\newrobustcmd{\tr}{\underset{\res}{\Tr}}
\newrobustcmd{\trd}{\underset{\text{D}}{\Tr}}
\newrobustcmd{\tri}{\Tr_\res}
\newrobustcmd{\col}[4]{{\begin{bmatrix}#1 \\ #2 \\ #3 \\ #4 \end{bmatrix}}}
\newrobustcmd{\row}[4]{{\begin{bmatrix}#1 &  #2  & #3  & #4 \end{bmatrix}}}
\newrobustcmd{\suppmat}{\cite{Schulenborg15Suppmat}}
\newrobustcmd{\refsuppmat}{\cite{Schulenborg15Suppmat}}
\newrobustcmd{\Eq}[1]{Eq.~(\ref{#1})}
\newrobustcmd{\eq}[1]{(\ref{#1})}
\newrobustcmd{\Fig}[1]{Fig.~\ref{#1}}
\newrobustcmd{\fig}[1]{\ref{#1}}
\newrobustcmd{\Figs}[1]{Figs.~\ref{#1}}
\newrobustcmd{\Sec}[1]{Sec.~\ref{#1}}
\newrobustcmd{\Ref}[1]{Ref.~[\onlinecite{#1}]}
\newrobustcmd{\Refs}[1]{Refs.~[\onlinecite{#1}]}
\newrobustcmd{\RefsSupp}[1]{Refs.~\cite{#1}}
\definecolor{grey}{rgb}{0.75,0.75,0.75}
\definecolor{orange}{rgb}{1.0,0.5,0.5}
\definecolor{brown}{rgb}{0.5,0.25,0.0}
\definecolor{pink}{rgb}{1.0,0.4,0.0}
\definecolor{green}{rgb}{0.0,0.75,0.0}
\definecolor{darkblue}{rgb}{0.0,0.0,0.75}
\definecolor{red}{rgb}{1.0,0.0,0.0}
\definecolor{darkred}{rgb}{1.0,0.0,0.0}
\newrobustcmd{\todo}[1]{{\color{red}TODO: #1}}
\newrobustcmd{\cut}[1]{}
\newrobustcmd{\Cutl}[2]{}
\newrobustcmd{\Cutr}[2]{}
\newrobustcmd{\com}[1]{}
\newrobustcmd{\new}[1]{{#1}}
\newrobustcmd{\New}[2]{{#1}}
\newrobustcmd{\Newl}[2]{{#1}}
\newrobustcmd{\Newr}[2]{{#1}}
\newrobustcmd{\advances}[1]{#1}
\newrobustcmd{\opens}[1]{#1}
\newrobustcmd{\outstanding}[1]{#1}
\newrobustcmd{\general}[1]{#1}
\newrobustcmd{\self}[1]{#1}
\newrobustcmd{\change}[1]{#1}

\newrobustcmd{\red}[1]{\textcolor{red}{#1}}

\usepackage{placeins}

\makeatother

\begin{document}

\title{Spin switching via quantum dot spin valves}

\author{N. M. Gergs}

\affiliation{Institute for Theoretical Physics, Center for Extreme Matter and
Emergent Phenomena, Utrecht University, Leuvenlaan 4, 3584 CE Utrecht,
The Netherlands}

\author{S. A. Bender}

\affiliation{Institute for Theoretical Physics, Center for Extreme Matter and
Emergent Phenomena, Utrecht University, Leuvenlaan 4, 3584 CE Utrecht,
The Netherlands}

\author{R. A. Duine}

\affiliation{Institute for Theoretical Physics, Center for Extreme Matter and
Emergent Phenomena, Utrecht University, Leuvenlaan 4, 3584 CE Utrecht,
The Netherlands}

\affiliation{Department of Applied Physics, Eindhoven University of Technology,
5600 MB, Eindhoven, The Netherlands}

\author{D. Schuricht}

\affiliation{Institute for Theoretical Physics, Center for Extreme Matter and
Emergent Phenomena, Utrecht University, Leuvenlaan 4, 3584 CE Utrecht,
The Netherlands}

\date{\today}
\begin{abstract}
We develop a theory for spin  transport and magnetization dynamics in a quantum-dot spin valve, i.e., two magnetic reservoirs coupled to a quantum dot. Our theory is able to take into account effects of strong correlations. We demonstrate that, as a result of these strong correlations, the dot gate voltage enables control over the current-induced torques on the magnets, and, in particular, enables voltage-controlled  magnetic switching. The electrical resistance of the structure can be used to read out the magnetic state. Our model may be realized by a number of experimental systems, including magnetic scanning-tunneling microscope tips and artificial quantum dot systems. 
\end{abstract}


\maketitle
\pagestyle{plain}
\maketitle

\global\long\def\ii{\text{i}}

\emph{Introduction}.---The reliable manipulation and detection of magnetic moments by electrical means remains one of the overarching themes of spintronics. Recent years have seen the development of several techniques involving a variety of materials (conducting, insulating and semiconducting) and heterostructures to this end. A key observation is that the total conductance of metallic magnetic multilayers may be extremely sensitive to the magnetic orientations of the constituent magnets~\citep{Baibich88,Camley89,Levy90,Valet93}, owing to the spin-dependent transport coefficients of the various components; the ``giant magnetoresistance" of such heterostructures demonstrates the possibility of electrically reading the magnetic state of microlayers and has been employed in mass produced devices shortly thereafter. Subsequently,  it was shown that the generation of magnetic dynamics leads to switching of magnetic multilayers by large electrical currents, which become spin polarized and thus transfer spin across the structure~\citep{Berger96,Tsoi98,Slonczewski99,Zhitao07,Ralph08,Evarts09}.

Typically, components of such heterostructures are sufficiently large that interactions and quantum effects do not play a prominent role in transport. As devices are scaled down, however, these effects become increasingly significant. Quantum dots coupled to ferromagnetic leads, which can be viewed as a nanoscale analogue to magnetic multilayer spin valves, represent an extreme scaling down of the metallic interlayer. These quantum dot spin valves have proven a fertile subject of research on spin-dependent quantum transport in recent years~\citep{Loss02,PRLKoenigMartinek,Martinek03,TransportTheoryKoenigMartinek,Braun05,Wilczynski05,Swirkowicz08,Merchant08,Hell2015}. In these studies, the ferromagnetic leads are static reservoirs of angular momentum; if, however, the reservoir magnetic moments are sufficiently small and the electric currents sufficiently large, the reservoir moments may be reoriented by the absorption of spin current (i.e., spin-transfer torque), just as in magnetic multilayers. In contrast to multilayers, however, wherein spin-transfer torques are controlled by the source-drain bias alone, in ferromagnet-dot-ferromagnet tunnel junctions, gating of the dot provides a new route of electrical manipulation of magnetic dynamics, opening up rich new phase behavior for the magnetic orientations of the ferromagnetic reservoirs.

In this Letter we discuss how transport through a spin-degenerate quantum dot can be utilized to manipulate attached nanomagnets via applied gate and bias voltages. These voltages control the electronic transport through the quantum dot, which in turn induces spin torques in the nanomagnets. Since the electronic transport can be well controlled by the gate and bias voltage, the spin torques can be tuned as well, which is not straightforwardly possible in standard spintronics setups like magnetic multilayers. We demonstrate that the tunablility of the spin torques enables the magnetic switching of the nanomagnets between parallel and anti-parallel configurations, which are experimentally distinguishable by their magnetoresistance and thus can be readout electrically. Moreover, we find a new resonance for nearly parallelly aligned reservoirs which turns out to be a partner resonance to the recently reported~\citep{Hell2015} spin resonance in the absence of spin splitting.

We suppose a separation of timescales between ``fast'' quantum electron transport and ``slow'' magnetic dynamics of the reservoirs. While such an approach assumes low magnetic frequencies, and thus that charge- and spin-pumping effects are negligible, it allows us to first treat electron transport for quasistatic, arbitrary magnetic orientations of the reservoirs. We then use the resulting expressions for spin-polarized current that flows through the structure to obtain spin torques on the magnetic reservoirs, which drive magnetic dynamics. This article thus combines two approaches. The first is a semiclassical treatment of the spin torques and ``slow'' magnetic dynamics from within a Landau--Lifshitz--Gilbert (LLG) phenomenology. The second deals with the ``fast'' electron transport through the quantum dot/quasistatic magnetic reservoirs; the quantum dot spin valve transport properties we obtain are quite general and apply to a wide range of magnetic systems in which a quantum dot might be embedded.

\emph{Magnetic dynamics}.---To model magnetic dynamics of the reservoirs, induced by electronic transport discussed below, we treat their respective magnetic moments as single-domain macrospins subject to the LLG equations modified to incorporate spin torques~\citep{Slonczewski96}:
\begin{equation}
\change{
S\frac{d\vec{n}_{r}}{dt}=-\mu_0\gamma S\vec{n}_{r}\times\vec{H}_{r}+\vec{I}_{\textrm{S},r}^{\bot}-\alpha S\vec{n}_{r}\times\frac{d\vec{n}_{r}}{dt},}
\label{eq:LLG}
\end{equation}
where $S$ is the macrospin of the reservoirs, $r\in\left\{ \textrm{S},\textrm{D}\right\}$ denotes the source and drain with the macrospin orientations $\vec{n}_{r}$ ($|\vec{n}_{r}|=1$), $\gamma$ is the absolute value of the gyromagnetic ratio, \change{$\mu_0$ the vacuum permeability,} and $\alpha$ is the phenomenological Gilbert-damping. The effective magnetic field in the reservoirs is given by \change{$\vec{H}_{r}=\frac{1}{\mu_{0}\gamma S}\frac{\delta E_{r}}{\delta\vec{n}_r}+\vec{H}_{\textrm{thermal}}$.} We consider the simple case of an easy-axis energy $E_{r}=-\frac{KV}{2}\left(\vec{n}_r\cdot\vec{e}_{\textrm{z}}\right)^{2}$, 
which facilitates two degenerate magnetic equilibria at $n_{r}^z=\pm1$. 
Motivated by materials like Galfenol~\citep{ExpValve1,ExpValve3} (iron-gallium alloys) we have also analyzed~\cite{supplement} the case of cubic anisotropy and found that all qualitative features remain unchanged. 
Furthermore, $\vec{H}_{\textrm{thermal}}$ implements the influence of the temperature $T$ via fluctuations with Gaussian noise of variance~\citep{LLGThermal} $\sigma_{\textrm{thermal}}^{2}=\frac{2\alpha T}{\gamma^2\mu_0^2 S}$.
Finally, $\vec{I}_{\textrm{S},r}^{\bot}$ is the component of the electronic spin current $\vec{I}_{\textrm{S},r}$ impinging on the reservoirs that is perpendicular to the macrospin orientation $\vec{n}_{r}$. Whereas the parallel component is carried into the bulk of the reservoir by itinerant electrons, $\vec{I}_{\textrm{S},r}^{\bot}$ is absorbed by reorienting $\vec{n}_{r}$, and hence enters as a spin torque~\citep{Brataas08}; this current, which gives rise to magnetic dynamics and switching, is highly sensitive to interactions in the magnetic dot and will be calculated and discussed in detail in the next subsection. The spin current can be further decomposed~\citep{Slonczewski96} into an out-of-plane spin current $I_{\textrm{FL},r}$ that acts field-like in the LLG-equation \eqref{eq:LLG} and an in-plane damping-like contribution $I_{\textrm{DL},r}$,
\begin{equation}
\vec{I}_{\textrm{S},r}^{\bot}:=\left(\vec{n}_{r}\times\vec{e}_{\textrm{z}}\right)I_{\textrm{FL},r}+\vec{n}_{r}\times\left(\vec{n}_{r}\times\vec{e}_{\textrm{z}}\right)I_{\textrm{DL},r}.\label{eq:torque}
\end{equation}
For simplicity we will limit our discussion to the case where the magnetization direction of only the drain is able to rotate freely, with the source macrospin $\vec{n}_{\textrm{S}}$ fixed at $\vec{e}_{\textrm{z}}$. For the case that both nanomagnets are able to rotate freely, no qualitative changes occur inside the Coulomb blockade regime where the quantum dot is singly occupied. 

The perpendicular spin current $\vec{I}_{\textrm{S,r}}^{\bot}$ in Eq.(\ref{eq:LLG}) includes all effects of the coupling of the two nanomagnets via the quantum dot device. In the next section, we discuss the quantum dot, the electronic transport, and the resulting spin current $\vec{I}_{\textrm{S},r}$.

\begin{figure}
\begin{centering}
\includegraphics[width=1\columnwidth]{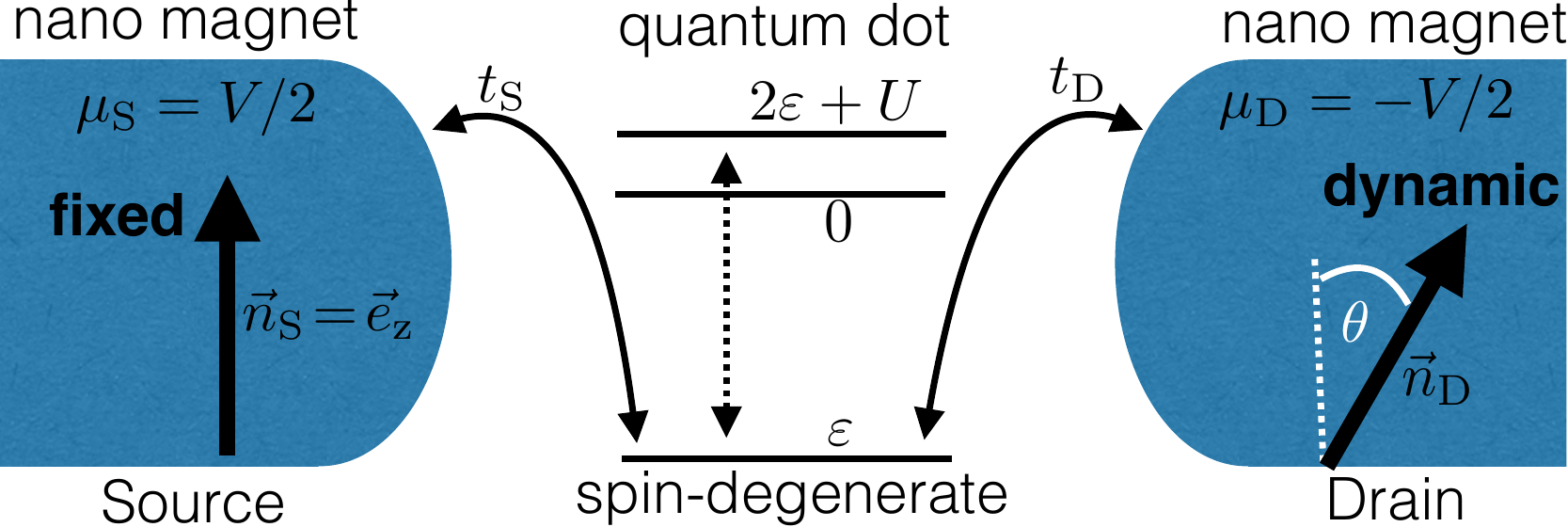}
\par\end{centering}
\caption{Sketch of the system: A spin degenerate single-level quantum dot is coupled via tunnel couplings $t_\text{S/D}$ to ferromagnetic source and drain reservoirs, which are held at different chemical potentials $\mu_\text{S/D}=\pm V/2$. The orientation $\vec{n}_\text{S}$ of the source reservoir is fixed while the macrospin $\vec{n}_\text{D}$ of the drain can fluctuate.\label{fig:setup}}
\end{figure}

\emph{Transport}.---We consider as a simple model an Anderson impurit\textcolor{black}{y} with a spin-degenerate energy level coupled to generally non-collinear magnetic reservoirs. All qualitative conclusions are quite general and expected to hold also for systems with, for example, multiple energy levels or weak to moderate electron-phonon coupling as none of these fundamentally change the equation of motion for the dot spin. The dot and tunneling Hamiltonians are given by $H_{\textrm{dot}}=\sum_{\sigma}\varepsilon n_{\sigma}+Un_{\uparrow}n_{\downarrow}$, $n_\sigma=d_\sigma^\dagger d_\sigma$, $\sigma=\uparrow,\downarrow$ and $H_{\textrm{tun}}^{r}=\sum_{\sigma\sigma'k}t_{r}d_{\sigma'}^{\dagger}R_{\sigma'\sigma}^{r}c_{\sigma kr}+\textrm{h.c.}$, $ r\in\textrm{\{S,D\}}$, 
with $d_\sigma$ and $c_{\sigma kr}$ denoting the electron annihilation operators on the dot and in the reservoirs, and $t_{r}$ is the corresponding tunnel coupling. A sketch of the setup is given in Fig.~\ref{fig:setup}. In the following we will use the notation $\tilde{\varepsilon}:=\varepsilon+U/2$ which is the particle-hole-symmetrized \citep{Saptsov12a} on-site energy. In both reservoirs we consider a polarization implemented via the density of states $\rho^\sigma$ of spin $\sigma$, i.e., we set $p:=(\rho^{\uparrow}-\rho^{\downarrow})/(\rho^{\uparrow}+\rho^{\downarrow})$ and $\rho_0=(\rho^\uparrow+\rho^\downarrow)/2$. Furthermore, $R^{r}$ is a matrix that encodes the different polarization direction of the reservoirs. Without loss of generality, we choose for the source $R_{\sigma\sigma'}^{\textrm{S}}=\delta_{\sigma\sigma'}$ and hence for an opening angle $\theta$ between both reservoir polarization directions we obtain $R_{\sigma\sigma'}^{\textrm{D}}=\delta_{\sigma\sigma'}\cos\left(\frac{\theta}{2}\right)+\sigma\left(1-\delta_{\sigma\sigma'}\right)\sin\left(\frac{\theta}{2}\right)$.
The transport properties of this model have already been studied~\citep{PRLKoenigMartinek,TransportTheoryKoenigMartinek,Muralidharan13,Hell2015,Hoffman15} in some detail. In contrast, here we focus on the spin torques and in particular their consequences for the dynamics of the reservoir nanomagnets. 

To compute the transport through this strongly correlated quantum dot, we treat dot-reservoir tunnel coupling perturbatively~\citep{Leijnse08a,Schoeller09a,Leijnse09a,Koller10,Saptsov12a,Saptsov14a,supplement}. The corresponding perturbative scale is given by $\Gamma=\frac{1}{2}\sum_{r}\Gamma_{r}$ with $\Gamma_{r}:=2\pi t_{r}^{2}\rho_0$, i.e., we assume $\Gamma\ll T$. All simulations are carried out with a stepsize of $\Delta t=10^{-2}\Gamma$. We assume a flat density of states for the reservoirs with an energy cutoff $D$ much larger than all other energy scales in the model. We include all $\mathcal{O}\left(\Gamma\right)$ and $\mathcal{O}\left(\Gamma^{2}\right)$-effects like e.g. cotunneling in a consistent way regarding the occurring relaxation rates of the density operator. From this we determine the stationary transport properties of the model. The underlying assumption behind using the stationary values is that the timescale for the dot-relaxation processes to happen is much smaller than the timescale for the resulting dynamics of the nanomagnets described by \eqref{eq:LLG}.

Albeit we consider a simple model for the dot Hamiltonian, the non-spin-conserving tunnel couplings to the reservoirs yield a variety of spin phenomena to explore. In the equations of motion describing the dot dynamics, the finite reservoir polarizations yield an induced magnetic field on the dot given by~\citep{TransportTheoryKoenigMartinek}
\begin{equation}
\vec{B}_{\textrm{ind}}=\frac{1}{\pi}\sum_{\substack{r=\textrm{S},\textrm{D}\\
q=\pm
}
}\Gamma_{r}\vec{n}_{r}p_{r}q\,\textrm{Re}\,\psi\left(\frac{1}{2}+i\frac{\tilde{\varepsilon}-\mu_{r}-qU/2}{2\pi T_{r}}\right)\label{eq:effective-field}
\end{equation}
with $\psi$ denoting the digamma function. The induced effective field \eqref{eq:effective-field} is included in the non-vanishing leading order~\citep{supplement}. We note that the \change{near} degeneracy of the dot level is crucial as a large Zeeman splitting would pin the dot polarization direction and suppress the effects associated to the induced magnetic field. Such a degeneracy can also be realized in the presence of strong external magnetic fields by, for example, fine tuning~\citep{Brooks} different valley degrees of freedom. 

\emph{Spin resonance condition}.---Crucial for the understanding of our results is the fact that the  degenerate spin-level of the quantum dot leads, in combination with the induced effective magnetic field \eqref{eq:effective-field} from the reservoirs, to a non-trivial spin resonance~\citep{Hell2015}. The resonance condition relevant for the dynamics of the nanomagnets is given by (generalizing the result of Ref.~\onlinecite{Hell2015})
\begin{equation}
\vec{B}_{\textrm{ind}}\cdot\left(\vec{n}_{\textrm{D}}-\vec{n}_{\textrm{S}}\right)=0,\label{eq:res1}
\end{equation}
which is the condition for the effective magnetic field \eqref{eq:effective-field} to change the dot spin from one initial reservoir polarization direction to the other with maximal efficiency, \change{i.e., fastest in time and with the smallest necessary thermal fluctuations}. The resonance condition \eqref{eq:res1} \change{results in a relation between the bias voltage $V$ and on-site energy $\tilde{\varepsilon}$, which is independent of the opening angle, provided that $\theta\neq0$. For a symmetric setup, $p_\text{S}\Gamma_{\textrm{D}}=p_\text{D}\Gamma_{\textrm{S}}$,} the resonance condition reduces to $V=0$ for which transport is suppressed~\citep{Hell2015}.

\begin{figure}
\begin{centering}
\includegraphics[width=1\columnwidth]{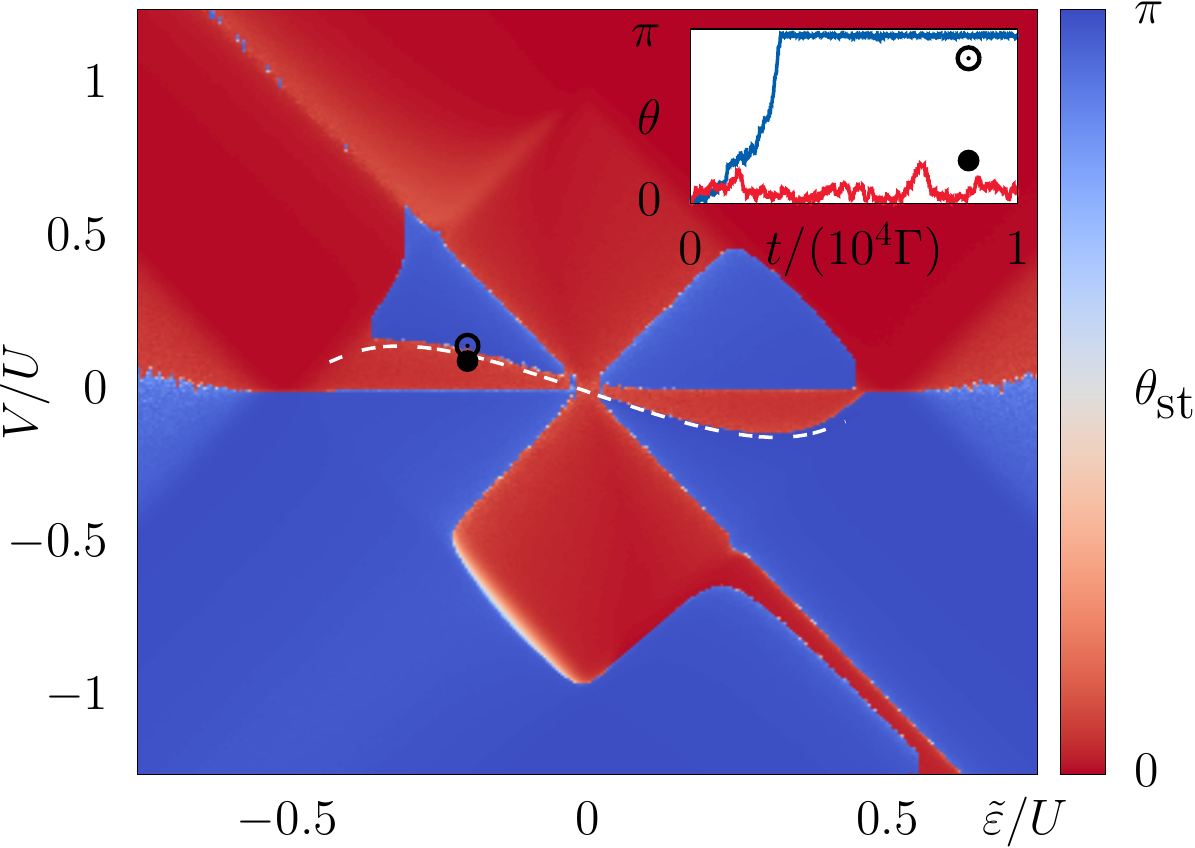}
\par\end{centering}
\caption{Switching diagram displaying the average opening angle between the reservoirs (see text for the precise definition). The parameters are given by $\Gamma_{\textrm{D}}=2\Gamma_{\textrm{S}}$, $U=10^2\,T=10^{3}\,\Gamma$, $p=0.99$, $KV=0.01\,\Gamma$, $\alpha=0.1$, and $\sigma_{\textrm{thermal}}^{2}=10^{-4}$. The resonance \eqref{eq:res1} is indicated by the white dashed line. Inset: Two exemplary time evolutions of the opening angle for the parameters $\tilde{\varepsilon}=0.2\,U$ and $V=0.1\,U$ as well as $V=0.175\,U$.} 
\label{fig:switching}
\end{figure}

\emph{Switching diagram}.---We will now turn to our main result and discuss the spin dynamics. For already small deviations from the parallel alignment $\theta\gtrsim0.1\pi$ the field-like spin torque dominates over the damping-like contribution; see Eq.~\eqref{eq:torque} for the definitions. Depending on the sign of the field-like torque, the driven magnetic ground state will be either a parallel or an anti-parallel alignment of the nanomagnets; we refer to these as the parallel and anti-parallel phases, respectively.

In Fig.~\ref{fig:switching} we show the corresponding phase diagram \change{(see also ~\cite{supplement})}. The plotted average opening angle between the source and drain reservoirs is obtained as follows: The system is initialized in a parallel configuration $\theta_\text{init}=0$ (recall that the orientation of the source spin is held fixed). The finite temperature encoded in the variance $\sigma_\text{thermal}$ leads to fluctuations in $\theta$ which are, depending on the point in the $(\tilde{\varepsilon},V)$-phase diagram, suppressed or enhanced by the spin \change{torque 
acting} on $\vec{n}_\text{D}$; two exemplary time evolutions of the angle $\theta$ are shown in the inset. \change{(We note that the macrospin length $S$ only rescales the time axis.)} The average opening angle is obtained by performing the \change{time evolution up to $t=10^4\,S/\Gamma$ and then averaging $\theta$ from there to $t=2\cdot 10^4\,S/\Gamma$.} In the phase diagram we see that the average opening angle indeed ends up in either a parallel or an anti-parallel configuration, and that this can be controlled by the values of the gate and bias voltages $\varepsilon=\tilde{\varepsilon}-U/2$ and $V$ respectively. For example, the resonance \eqref{eq:res1} indicated by the white dashed line corresponds to a phase boundary. The precise value of the stationary angle depends on the parameters, in particular also on the temperature. We note that the lack of symmetry of the switching diagram under $\left(\tilde{\varepsilon},V\right)\to\left(-\tilde{\varepsilon},-V\right)$ is due to the fixing of the source magnetization $\vec{n}_\text{S}$. 

\begin{figure}
\begin{centering}
\includegraphics[width=1\columnwidth]{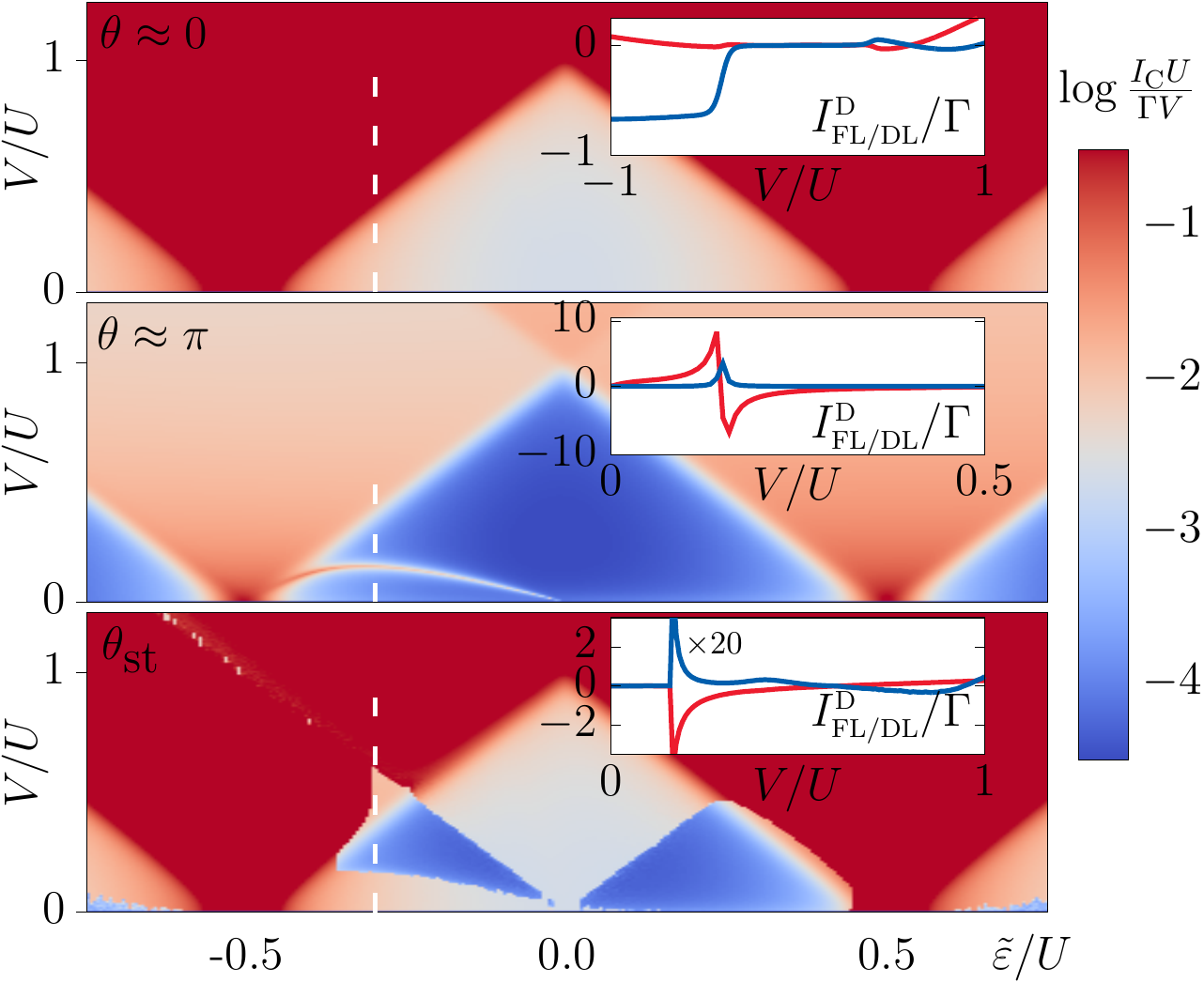}
\par\end{centering}
\caption{Logarithmic representation of the charge current per applied voltage bias $\log(I_{\textrm{C}}/V)$ in the nearly parallel ($\theta=0.01\pi$), nearly anti-parallel ($\theta=0.99\pi$), and dynamic cases (from Fig.~\ref{fig:switching}). All parameters are as in Fig.~\ref{fig:switching}. The upper part of the color scale relevant for transport outside of the Coulomb blockade is omitted. Generically the parallel alignment results in larger currents than the anti-parallel alignment. Insets: Field-like (red) and damping-like (blue) spin torques along cuts at $\tilde{\varepsilon}=-0.3\,U$ indicated by dashed while lines in the main figure.\label{fig:readout}}
\end{figure}

A transition between parallel and anti-parallel phases can be easily detected by measuring the charge transport through the device. Let us compare three cases: one in which the drain magnetization is fixed in the parallel configuration ($\theta\approx 0$), one in which it is fixed in the anti-parallel configuration ($\theta\approx\pi$), and one in which the relative orientation $\theta_{\mathrm{st}}$ is determined by the applied voltages as shown in the phase diagram Fig.~\ref{fig:switching}. As shown in Fig.~\ref{fig:readout}, charge transport is qualitatively distinct between the three cases. For example, the resonance \eqref{eq:res1} does not~\citep{Hell2015} cause a significant feature in the parallel case, while in the other two cases it is clearly visible in the charge transport. The insets in Fig.~\ref{fig:readout} show the field-like (red lines) and damping-like (blue lines) spin torques along cuts at $\tilde{\varepsilon}=-0.3\,U$, revealing nontrivial dependencies on the applied gate and bias voltages. In particular we conclude that the quantum-dot setup allows control over the relative strength of the spin torques. 

\begin{figure}
\begin{centering}
\includegraphics[width=1\columnwidth]{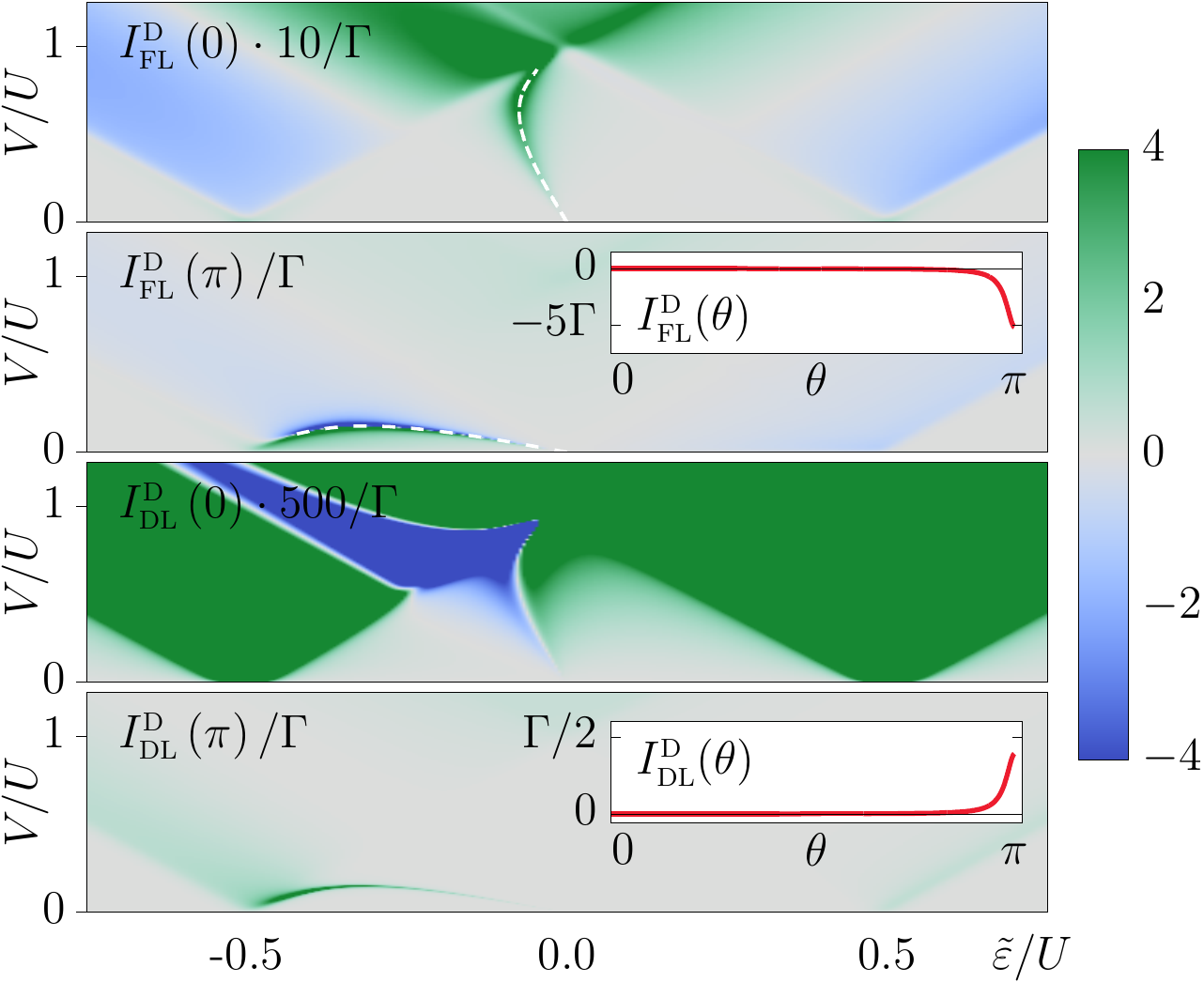}
\par\end{centering}
\caption{Field- and damping-like spin torques $I_{\textrm{FL}}^{\textrm{D}}$ and $I_{\textrm{DL}}^{\textrm{D}}$ for the parallel (respective upper) and anti-parallel (respective lower) cases. All parameters are as in Fig.~\ref{fig:switching}. The dashed lines correspond to the resonance conditions $\vec{B}_{\textrm{ind}}\cdot(\vec{n}_{\textrm{D}}\pm\vec{n}_{\textrm{S}})=0$, with the upper and lower sign valid for $\theta\approx 0$ and $\theta\approx\pi$ respectively. The full phase diagrams are obtained from the symmetry and anti-symmetry of $I_{\textrm{FL}}^{\textrm{D}}$ and $I_{\textrm{DL}}^{\textrm{D}}$ under $(\tilde{\varepsilon},V)\to(-\tilde{\varepsilon},-V)$. Insets: Angular dependence of the respective spin torque for \change{$\tilde{\varepsilon}=-U/4$, $V=0.15\,U$} close to the resonance \eqref{eq:res1}.\label{fig:FMR}}
\end{figure}

\change{The actual switching from anti-parallel to parallel always occurs via a fast transition, while the inverse process is significantly influenced by temperature. The underlying reason for this is the absence of spin torques in the parallel limit (see insets in Fig.~\ref{fig:FMR}). Hence, thermal fluctuations first have to generate a sufficiently large initial opening angle for the field-like spin torque to take over and drive the actual switching. As this waiting time is usually much longer than the actual spin-torque driven switching time~\citep{supplement}, this effect dominates. On the other hand, due to the stronger spin torques the anti-parallel alignment---as long as it is supported by the spin current---is highly robust against thermal fluctuations, counterintuitively in particular for $(\varepsilon,V)$-values close to the resonance where the transition from anti-parallel to parallel alignment occurs. In contrast, going across the transition line in the $(\varepsilon,V)$-parameter regime will cause similar strong spin currents to easily switch the nanomagnets to a parallel alignment.}

\emph{FMR}.---We now turn to the spin torques shown in Fig.~\ref{fig:FMR}. These are accessible via ferromagnetic resonance (FMR) experiments. We assume that the corresponding external magnetic field for such a setup is small enough to not effectively lift the degeneracy of the quantum dot, i.e., $B_{\textrm{extern}}\ll\Gamma$. For larger external magnetic field the precessing quantum dot spin would be pinned and the effects discussed here will be suppressed.

FMR measures the absorption of electromagnetic waves at a given frequency $\Omega$.  The response of the $x$-component of the drain spin density to, for example, a linearly polarized driving field $\mathbf{h}(t)=h_x \mathrm{cos} \left( \Omega t \right) \hat{\mathbf{x}}$ is given by $n_{\text{D},x}(t)=\chi'_{xx} h_x \mathrm{cos}(\Omega t)+\chi''_{xx} h_x \mathrm{sin}(\Omega t)$. Assuming $\alpha \ll 1$, the out-of-phase susceptibility $\chi''_{xx}$ is, for example~\citep{supplement}
\begin{equation}
\chi''_{xx}=S\Omega \frac{-2S^{2}\Omega_{R}^2\alpha_{\mathrm{eff}}+\alpha \left[\left(I_{\mathrm{DL}}^{\mathrm{D}}\right)^{2}+S^{2}\left(\Omega_{R}^{2}-\Omega^{2}\right)\right]}{\left[\left(I_{\mathrm{DL}}^{\mathrm{D}}\right)^{2}+S^{2}\left(\Omega_{R}^{2}-\Omega^{2}\right)\right]^{2}+\left(2S^{2}\alpha_{\mathrm{eff}}\Omega\Omega_{R}\right)^{2}}
\end{equation}
where $\Omega_{R}=\left(\gamma H-I_{\textrm{FL}}^\text{D}\right)/S$ and $\alpha_{\textrm{eff}}=\alpha+I_{\textrm{DL}}^\text{D}/(S\Omega_R)$. Hence, the field-like torque causes a shift of the resonance position, while the damping-like torque alters the broadening $\sim \alpha_{\mathrm{eff}}$. 

As shown in Fig.~\ref{fig:FMR}, for an anti-parallel configuration, $\theta\approx\pi$, the field- and damping-like torques are of the same order and both clearly exhibit a feature following the resonance condition \eqref{eq:res1}. On the other hand, for the parallel configuration the damping-like spin torque inside the Coulomb blockade is heavily suppressed. Furthermore, instead of the resonance at \eqref{eq:res1} both the field- and daming-like torque possess a new feature at $\vec{B}_{\textrm{ind}}\cdot(\vec{n}_{\textrm{D}}+\vec{n}_{\textrm{S}})=0$. This resonance is the partner of \eqref{eq:res1} in the sense that, instead of maximizing, it minimizes the efficiency of the induced magnetic field to rotate the spin from the source to the drain direction or vice versa. 
It also appears in the cotunneling charge current through the quantum dot and yields a very weak conductance peak (invisible in Fig.~\ref{fig:readout}). 
If the opening angles start to have a relevant deviation from the nearly-parallel setup $\theta\gtrsim0.1\pi$ the resonance following \eqref{eq:res1} quickly begins to dominate.

\emph{Controlling the spin torques}.---Generally the strong dependence of the field- and damping-like spin torques on the tunable gate voltage $\varepsilon$ (via the on-site energy $\tilde{\varepsilon}$) and the applied voltage bias $V$ allow for experimental control of these normally fixed parameters. The insets of Fig.~\ref{fig:readout} show how the spin torques vary as a function of the bias voltage $V$, corresponding to cuts in Fig.~\ref{fig:FMR}. For example, in the fixed parallel case one can switch on the damping-like torque by leaving the Coulomb diamond. In contrast, the fixed anti-parallel setup has significant features only inside the Coulomb blockade region where around the resonance at \eqref{eq:res1} the sign of the field-like torque can be flipped. 

\emph{Conclusions}.---We showed that quantum transport through a spin-degenerate quantum dot provides unique control over the spin torques acting on the attached nanomagnets, thus enabling the effective switching of the nanomagnets from a parallel to an anti-parallel configuration and vice versa. At the same time, the cotunneling charge current through the system is a reliable readout tool for the magnetic configuration of the nanomagnets. Our finding relies on the spin-degeneracy of the quantum dot level; adding further aspects to the model like additional quantum levels or phonon degrees of freedom is not expected to change our results qualitatively. 

We would like to thank Gerrit Bauer, Michael Hell, Martin Leijnse, Yaroslav Tserkovnyak, and Maarten Wegewijs  for useful comments and discussions. This work is part of the D-ITP consortium, a program of the Netherlands Organisation for Scientific Research (NWO) that is funded by the Dutch Ministry of Education, Culture and Science (OCW). This project has received funding from the European Research Council (ERC) under the European Union's Horizon 2020 research and innovation programme (grant agreement 725509 - SPINBEYOND).

\newpage
\phantom{O}
\newpage
\onecolumngrid
\setcounter{figure}{0}

\begin{center}
\textbf{\large Supplement: Spin switching via quantum dot spin valves}
\end{center}

\section{Real-time perturbation theory}

\subsection{General considerations}

Here we will give a brief recap of the real-time perturbation theory~\citep{Schoeller09a,Schoeller09bA,Saptsov12a,Leijnse08a}
used for the quantum transport computation in the main part of this
letter. Most of this recap will focus on how to compute the decay
rates of the quantum dot system under consideration. In subsection
\ref{sub:Charge-and-spin} we will comment on the computation of the
charge and the spin currents as they require the theory build around
the decay rates as prerequisite.

The main assumption of this method is that the Hamiltonian can be
factored in three parts $H=H_{\textrm{dot}}+H_{\textrm{res}}+H_{\textrm{tun}}$.
The quantum dot Hamiltonian $H_{\textrm{dot}}$ describes the system
of interest, which is typically small (i.e. possesses a low-dimensional Hilbert
space) and will be treated exactly throughout this method. The reservoir
Hamiltonian usually describes macroscopic, non-interacting reservoirs
that are e.g. given by 
\begin{equation}
H_{\textrm{res}}=\sum_{\substack{k,\sigma=\uparrow,\downarrow\\
r=\textrm{S},\textrm{D}
}
}\omega_{k}c_{kr\sigma}^{\dagger}c_{k\sigma}\label{eq:Hamiltonian-res}
\end{equation}
and are assumed to be at grandcanonical equilibrium. The crucial assumption
for this perturbative approach is now that the coupling between this
quantum systems, described by the tunnel Hamiltonian $H_{\textrm{tun}}$,
is weak compared to the temperature of the attached reservoirs. For
this discussion we will assume a rather general form with non-spin-conserving
tunnel couplings $t_{\sigma\sigma'}^{r}\in\mathbb{R}$:
\begin{equation}
H_{\textrm{tun}}=\sum_{\substack{k,\sigma,\sigma'=\uparrow,\downarrow\\
r=\textrm{S},\textrm{D}
}
}t_{\sigma\sigma'}^{r}c_{kr\sigma}^{\dagger}d_{\sigma'}+\textrm{h.c.}\,\textrm{.}\label{eq:Hamiltonian-tun}
\end{equation}
We will furthermore assume that the initial density operator of the system
can always be factorized with respect to the dot and reservoir subsystems,
i.e. $\rho=\rho_{\textrm{dot}}\otimes\rho_{\textrm{res}}\,\textrm{.}$
The perturbative condition is now given by 
\[
\max\limits _{\substack{\sigma,\sigma'=\uparrow,\downarrow\\
r=\textrm{S},\textrm{D}
}
}\left(t_{\sigma\sigma'}^{r}\right)^{2}\rho_0\ll T_{r}
\]
where $T^{r}$ is the temperature of the corresponding macroscopic
reservoir $r$ and $\rho_0$ its density of states.

We now start the actual derivation of the perturbative description
of the system dynamics by considering the Liouville-von-Neumann equation:
\begin{equation}
i\frac{\partial}{\partial t}\mbox{\ensuremath{\rho}}=\left[H,\rho\right]_{-}=L\rho\label{eq:von-Neumann-full}
\end{equation}
where $\rho$ is the density operator of the full system and the superoperator
$L\bullet:=\left[H,\bullet\right]_{-}$ describing the dynamics of
the density operator will be referred to as Liouvillian. With respect
to the quantum dot system we will only be interested in its stationary
properties. Hence, we Laplace transform the Liouville-von-Neumann
equation \eqref{eq:von-Neumann-full}. Furthermore, we trace out the
static macroscopic reservoir degrees of freedom to obtain:
\begin{align}
\rho_{\textrm{dot}}\left(z\right) & =\Pi\left(z\right)\rho_{\textrm{dot}}\left(t_{0}\right)\label{eq:effective-dot}\\
\Pi\left(z\right) & =\underset{{\scriptstyle \textrm{res}}}{\textrm{tr}}\frac{i}{z-L_{\text{tot}}}\rho_{\text{res}}\nonumber 
\end{align}
where $t_{0}$ is some arbitrary initial time. Now this is rewritten
further by using the Dyson equation combined with the property
that the reservoir creation-/annihilation-operator satisfy Wick's
theorem due to the grandcanonical description of the reservoirs.
\begin{align}
\Pi\left(z\right) & =\frac{i}{z-L_{\text{eff}}\left(z\right)}\label{eq:effective-dot2}\\
L_{\text{eff}}\left(z\right): & =L_{\text{dot}}+\sum_{k=1}^{\infty}\left.\underset{{\scriptstyle \textrm{res}}}{\textrm{tr}}\left(L_{\text{tun}}\frac{1}{z-\left(L_{\text{dot}}+L_{\text{res}}\right)}\right)^{k}L_{\text{tun}}\rho_{\text{res}}\right|_{\text{irred.}}\label{eq:liouville-series}
\end{align}
Here irreducibility refers to the diagrams occurring the perturbative
series, for the details of the diagrammatic rules see Refs.~\citep{Schoeller09a,Schoeller09bA,Saptsov12a,Leijnse08a}.
The effective Liouvillian $L_{\text{eff}}\left(z\right)$ describes
now  via \eqref{eq:effective-dot} and \eqref{eq:effective-dot2} the
time-evolution of the quantum dot subsystem. The influence of the
tunnel coupling to the reservoirs is fully encoded in $L_{\text{eff}}\left(z\right)$
and up to now this description is exact. 
For our perturbative treatment we will cut off the series \eqref{eq:liouville-series}
at $k=5$. This corresponds to including effects of the order $L_{\textrm{tun}}^{4}$,
i.e. cotunnelling effects. Considering the inverse of the Laplace
transformation we can obtain the stationary dot density
operator via:
\begin{equation}
L_{\text{eff}}\left(z=i0^{+}\right)\rho_{\textrm{dot, stat.}}=0.\label{eq:dot-density}
\end{equation}
The existence of this eigenvector of the effective Liouvillian $L_{\text{eff}}\left(z=i0^{+}\right)$
follows from the fact that the time-evolution conserves the trace
of any quantum system,
\[
\underset{{\scriptstyle \textrm{dot}}}{\textrm{tr}}L_{\text{eff}}\left(z=i0^{+}\right)=0,
\]
corresponds to the left eigenvector of the zero-eigenstate of $L_{\text{eff}}\left(z=i0^{+}\right)$
and the stationary dot density operator \eqref{eq:dot-density} to
the respective right eigenvector. We stress that no restriction
on the elements of the density operator is required or used at
any point for this approach.

\subsection{Concrete expressions}

We now give some concrete expressions for our system.
We will use a creation-/annihilation-index $\eta$ defined as
\[
d_{\eta\sigma}=\begin{cases}
d_{\sigma} & \textrm{for }\eta=-\\
d_{\sigma}^{\dagger} & \textrm{for }\eta=+
\end{cases}
\]
and a multi-index $1=\left(\eta_{1},\sigma_{1}\right)$. The same
definitions, with an additional reservoir index $r$ and momentum
index $k$, will also be used for the reservoir operator. To have
nice commutator relations, we define the following mapping for the
creation and annihilation operators
\begin{equation}
\mathcal{G}_{1}^{q_{1}}\bullet=\frac{1}{\sqrt{2}}\left(d_{1}\bullet+q_{1}\left(-\mathcal{I}\right)^{N}\bullet\left(-\mathcal{I}\right)^{N}d_{1}\right),\qquad q_{1}\in\left\{ +,-\right\}, \label{eq:our-supercreate}
\end{equation}
and analogous with the symbol $\mathcal{J}_{1}^{q_{1}}$ for the reservoir
creation/annihilation operators $c_{1}$. The commutation relations
are now given by $\left[\mathcal{G}_{2}^{q_{2}},\mathcal{G}_{1}^{q_{1}}\right]_{+}=\delta_{q_{2},\overline{q}_{1}}\delta_{2,\overline{1}}\mathcal{I}$.
For the reservoir-density operator Wick's theorem can now be formulated
as follows~\footnote{we only consider $n$ even as the expectation values vanish for $n$
odd.}:
\begin{align}
\left\langle \mathcal{J}_{n}^{q_{n}}\dots\mathcal{J}_{1}^{q_{1}}\right\rangle _{\text{res}} & =\sum_{P}\left(-1\right)^{P}\prod_{\left\langle j,i\right\rangle }\left\langle \mathcal{J}_{j}^{q_{j}}\mathcal{J}_{i}^{q_{i}}\right\rangle, \label{eq:Wick-theorem}\\
\left\langle \mathcal{J}_{2}^{-}\mathcal{J}_{1}^{-}\right\rangle _{\text{res}} & =\delta_{q_{2},-}\delta_{2,\overline{1}}\gamma_{1}^{q_{1}},\nonumber 
\end{align}
where $\gamma_{1}^{+}=1$ and $\gamma_{1}^{-}=\tanh\frac{\eta_{1}\left(\omega_{1}-\mu_{1}\right)}{2T_{1}}$
are the symmetric and antisymmetric part of the Fermi--Dirac distribution
with chemical potential $\mu_{1}$ and temperature $T_{1}$. The sign-factor
$\left(-1\right)^{P}$ counts how many permutations are required to
rearrange the superoperators $\mathcal{J}$ into the right-hand-side
expression of \eqref{eq:Wick-theorem}. In the representation \eqref{eq:our-supercreate}
the Liouvillians corresponding to the tunnel and reservoir Hamiltonians
\eqref{eq:Hamiltonian-res} and \eqref{eq:Hamiltonian-tun} , e.g. $L_{\textrm{tun}}\bullet=\left[H_{\textrm{tun}},\bullet\right]_{-}$,
take the form:
\begin{align}
L_{\text{tun}} & =\sum_{2}t_{2'2}\eta_{2}\sum_{q_{2}}\mathcal{G}_{2'}^{\overline{q}_{2}}\mathcal{J}_{\overline{2}}^{q_{2}},\label{eq:Liouvillians-tun}\\
L_{\text{res}} & =\sum_{2}\overline{\omega}_{2}\mathcal{J}_{2}^{+}\mathcal{J}_{\overline{2}}^{-},\label{eq:Liouvillians-res}
\end{align}
where $\overline{\omega}_{2}:=\eta_{2}\omega_{2}$ and $t_{2'2}:=\delta_{r_{2}r_{2}'}t_{\sigma_{2}\sigma_{2'}}^{r_{2}}$.
Substituting these expressions into the effective Liouvillian $L_{\textrm{eff}}\left(z\right)$
\eqref{eq:liouville-series} yields with $\left[L_{\text{res}},\mathcal{J}_{1}^{q}\right]_{-}=\overline{\omega}_{1}\mathcal{J}_{1}^{q}$
and after applying Wick's theorem \eqref{eq:Wick-theorem} a remaining
matrix algebra for the dot Hilbert space and energy integrals for
the reservoir degrees of freedom. We define a decay rate $\Gamma_{\sigma_{1}\sigma_{1'}}^{r_{1}}=2\pi\rho_{0}\sum_{\sigma_{2}}t_{\sigma_{2}\sigma_{1}}^{r_{1}}t_{\sigma_{2}\sigma_{1'}}^{r_{1}}\left(1+p_{r}\sigma_{2}\right)$
where $\rho_{0}=(\rho_{\uparrow}+\rho_{\downarrow})/2$ being the
average density of states of the reservoirs. The spin-dependent polarization
factor $p_{r}$ of the density of states has already been absorbed
into the decay rate $\Gamma_{\sigma_{1}\sigma_{1'}}^{r_{1}}$. For
the two leading orders taken into account here we obtain then {[}in
the following $\sum\!\!\!\!\!\!\int_{\,\,\,1q_{1}}$ means integration
over $\omega_{1}$ and summation over all other indices occurring
in $1=(\eta_{1},\sigma_{1},r_{1},\omega_{1})${]}:
\begin{align}
L_{\text{eff}}\left(z\right)= & L_{\text{dot}}+\sum\!\!\!\!\!\!\!\!\!\int\limits _{1q_{1}}\frac{\Gamma_{\sigma_{1'}\sigma_{1}}^{r_{1}}}{2\pi}\mathcal{G}_{\overline{1}'}^{+}\frac{q_{1}\gamma_{1}^{q_{1}}}{\overline{\omega}_{1}+z-L_{\text{dot}}}\mathcal{G}_{1}^{\overline{q}_{1}}\nonumber \\
 & +\sum\!\!\!\!\!\!\!\!\!\!\!\!\int\limits _{12q_{1}q_{2}}\frac{\Gamma_{\sigma_{1'}\sigma_{1}}^{r_{1}}\Gamma_{\sigma_{2'}\sigma_{2}}^{r_{2}}}{\left(2\pi\right)^{2}}\left(\mathcal{G}_{\overline{1}'}^{+}\frac{1}{\overline{\omega}_{1}+z-L_{\text{dot}}}\mathcal{G}_{\overline{2}'}^{+}-\mathcal{G}_{\overline{2}'}^{+}\frac{1}{\overline{\omega}_{2}+z-L_{\text{dot}}}\mathcal{G}_{\overline{1}'}^{+}\right)\frac{\overline{q}_{2}\gamma_{2}^{\overline{q}_{2}}}{\sum\limits _{i=1,2}\overline{\omega}_{i}+z-L_{\text{dot}}}\mathcal{G}_{2}^{q_{2}}\frac{\overline{q}_{1}\gamma_{1}^{\overline{q}_{1}}}{z-L_{\text{dot}}+\overline{\omega}_{1}}\mathcal{G}_{1}^{q_{1}}.\label{eq:full-Liouvillian}
\end{align}
The operators $\mathcal{G}$ as well as the dot Liouvillian
$L_{\textrm{dot}}$ can be just numerically implemented as matrices
for sufficiently small dot Hilbert spaces. The only challenging part
that remains now is to solve the occurring integrals. For this purpose
we assume a flat reservoir density of states given by 
\[
\rho_{\textrm{res},\sigma}\left(\omega\right)=\rho_{0}\left(1+p\sigma\right)\Theta(D-\left|\omega\right|)
\]
 where $\Theta$ is the Heaviside-step-function and $D$ some cutoff
much larger than any other energy scale in the problem. The leading
order contributions $\mathcal{O}(\Gamma)$ can be solved exactly and
yield with $\psi$ being the Digamma-function:
\begin{equation}
L_{\text{eff}}\left(z\right)=L_{\text{dot}}+\sum_{\eta_{1}\sigma_{1'}\sigma_{1}r_{1}}\left(\frac{\Gamma_{\sigma_{1'}\sigma_{1}}^{r_{1}}}{\pi}\cdot\mathcal{G}_{\overline{1}'}^{+}\left[\psi\left(\frac{1}{2}-i\frac{z-L_{\text{dot}}+\overline{\mu}_{1}}{2\pi T_{1}}\right)-\ln\frac{D}{2\pi T_{1}}\right]\mathcal{G}_{1}^{+}-i\frac{\Gamma_{\sigma_{1'}\sigma_{1}}^{r_{1}}}{2}\mathcal{G}_{\overline{1}'}^{+}\mathcal{G}_{1}^{-}\right)+\mathcal{O}\left(\Gamma^{2}\right).\label{eq:Leading order}
\end{equation}
The imaginary part of the $\mathcal{O}(\Gamma)$-corrections
corresponds to the decay rates while the real part implements the
effective magnetic field on the dot induced by the reservoir polarizations~\citep{PRLKoenigMartinek,TransportTheoryKoenigMartinek}
\begin{equation}
\vec{B}_{\textrm{ind}}=\frac{1}{\pi}\sum_{\substack{r=\textrm{S},\textrm{D}\\
q=\pm
}
}\Gamma_{r}\vec{n}_{r}p_{r}q\textrm{Re}\,\psi\left(\frac{1}{2}+i\frac{\tilde{\varepsilon}-\mu_{r}-qU/2}{2\pi T_{r}}\right)\label{eq:magnetic-field}
\end{equation}
where $\Gamma_{r}=\sum_{\sigma}\Gamma_{r\sigma}/2$. It is stressed
that the magnetic field is automatically implemented in this method
hidden via the real-valued part of the $\mathcal{O}(\Gamma$)-contribution
to $L_{\textrm{eff}}(z)$ \eqref{eq:von-Neumann-full} and does not
require any special treatment. Hence, the above expression \eqref{eq:magnetic-field}
is only given for illustrative purposes and never used in this form
for any actual computation.

The additional $\mathcal{O}(\Gamma^{2})$-contributions are more involved.
The expressions for the imaginary decay rates are given in the supplement
of Ref.~[\citenum{Gergs15}]. For the real parts one can repeat the
derivation described therein. Only one of the two integrals can be
solved analytically by closing the integral along the upper complex
plane and using the residue theorem. The remaining integral can
also be treated with the residue theorem, but the remaining sum of $\tanh$-poles
needs to be carried out numerically~\footnote{Eq. (27) from the supplement of Ref.~[\citenum{Gergs15}] does not
need to be altered and also describes the real part. Hence, only the
derivations of Eq. (28) and (29) from the supplement of Ref.~[\citenum{Gergs15}]
needs to be redone.}. For completeness we list the corresponding expressions here:
\begin{align*}
\int_{-D}^{D}\int_{-D}^{D}d\overline{\omega}_{1}d\overline{\omega}_{2}\frac{1}{\overline{\omega}_{1}+z_{3}}\frac{\gamma_{1}^{-}}{\sum\limits _{i=1,2}\overline{\omega}_{i}+z_{2}}\frac{1}{\overline{\omega}_{1}+z_{1}} & =-i2\pi\frac{\psi\left(\frac{1}{2}-i\frac{z_{3}+\overline{\mu}_{1}}{2\pi T_{1}}\right)-\psi\left(\frac{1}{2}-i\frac{z_{1}+\overline{\mu}_{1}}{2\pi T_{1}}\right)}{z_{3}-z_{1}}\\
\int_{-D}^{D}\int_{-D}^{D}d\overline{\omega}_{1}d\overline{\omega}_{2}\frac{1}{\overline{\omega}_{1}+z_{3}}\frac{\gamma_{2}^{-}\gamma_{1}^{-}}{\sum\limits _{i=1,2}\overline{\omega}_{i}+z_{2}}\frac{1}{\overline{\omega}_{1}+z_{1}} & =-8\pi iT_{1}\sum_{n=0}^{k_{D}:=\frac{D}{2\pi T_{2}}-\frac{1}{2}}\phi\left(\frac{1}{2}-i\frac{z_{2}+\overline{\mu}_{1}+\overline{\mu}_{2}}{2\pi T_{2}}+\frac{T_{1}}{T_{2}}\left(n+\frac{1}{2}\right)\right)\\
 & \qquad\qquad\qquad\qquad\qquad\times\frac{1}{2\pi T_{1}\left(n+\frac{1}{2}\right)-i\left(z_{3}+\overline{\mu}_{1}\right)}\frac{1}{2\pi T_{1}\left(n+\frac{1}{2}\right)-i\left(z_{1}+\overline{\mu}_{1}\right)}\\
\int_{-D}^{D}\int_{-D}^{D}d\overline{\omega}_{1}d\overline{\omega}_{2}\frac{1}{\overline{\omega}_{2}+z_{3}}\frac{\gamma_{2}^{-}\gamma_{1}^{-}}{\sum\limits _{i=1,2}\overline{\omega}_{i}+z_{2}}\frac{1}{\overline{\omega}_{1}+z_{1}} &= -8\pi iT_{1}\sum_{n=0}^{k_{D}:=\frac{D}{2\pi T_{2}}-\frac{1}{2}}\frac{\psi\left(\frac{1}{2}-i\frac{z_{2}+\overline{\mu}_{1}+\overline{\mu}_{2}}{2\pi T_{2}}+\frac{T_{1}}{T_{2}}\left(n+\frac{1}{2}\right)\right)-\psi\left(\frac{1}{2}-i\frac{z_{3}+\overline{\mu}_{2}}{2\pi T_{2}}\right)}{2\pi T_{1}\left(n+\frac{1}{2}\right)-i\left(z_{2}-z_{3}+\overline{\mu}_{1}\right)}\\
 & \qquad\qquad\qquad\qquad\qquad\qquad\times\frac{1}{2\pi T_{1}\left(n+\frac{1}{2}\right)-i\left(z_{1}+\overline{\mu}_{1}\right)}
\end{align*}
where $\phi\left(\frac{1}{2}-i\frac{z}{2\pi T_{1}}\right):=-\psi\left(\frac{1}{2}-i\frac{z}{2\pi T_{1}}\right)+\ln\frac{D}{2\pi T_{1}}$.

\subsection{Charge and spin current\label{sub:Charge-and-spin}}

The starting point for any observable without an explicit time-dependence
is given by
\[
\left\langle A\right\rangle \left(z\right)=\underset{{\scriptstyle \text{res}}}{\text{tr}}A\rho\left(z\right)=\frac{1}{2}\text{tr}\,\underset{{\scriptstyle \text{res}}}{\text{tr}}L^{A,+}\frac{i}{z-L_{\text{tot}}}\rho_{\textrm{dot}}\left(t_{0}\right)\rho_{\text{res}}
\]
where $L^{A,+}\bullet:=\left[A,\bullet\right]_{+}$. By repeating
the perturbative scheme outline before for the computation of the
decay rates---i.e. applying the Dyson equation, using Wick's theorem
and regrouping everything in terms of irreducible diagrams---we obtain:
\begin{align}
\left\langle A\right\rangle \left(z\right) & =\frac{1}{2}\text{tr}\Sigma_{A,+}\left(z\right)\Pi\left(z\right)\rho_{\textrm{dot}}\left(t_{0}\right)\nonumber \\
 & =\frac{1}{2}\text{tr}\Sigma_{A,+}\left(z\right)\rho_{\textrm{dot}}\left(z\right)\nonumber \\
\text{where }\Sigma_{A,+}\left(z\right) & =\sum_{k=0}^{\infty}\left.\underset{{\scriptstyle \textrm{res}}}{\textrm{tr}}L^{A,+}\frac{1}{z-\left(L_{\text{dot}}+L_{\text{res}}\right)}\left(L_{\text{tun}}\frac{1}{z-\left(L_{\text{dot}}+L_{\text{res}}\right)}\right)^{k}L_{\text{tun}}\rho_{\text{res}}\right|_{\text{irred.}}\label{eq:observable-kernel}
\end{align}
This observable kernel $\Sigma_{A,+}\left(z\right)$
corresponds therefore to the corrections to the effective Liouvillian
$L_{\textrm{eff}}\left(z\right)-L_{\textrm{dot}}$ from \eqref{eq:liouville-series},
but the leftmost tunnelling Liouvillian has been replaced by the anticommutator
Liouvillian $L^{A,+}$ of the respective considered observable. All
further derivations leading to Eqs. \eqref{eq:full-Liouvillian} and
\eqref{eq:Leading order} therefore also hold for the observable kernel;
only the leftmost tunnelling Liouvillian needs to be replaced. The
stationary contribution can again be extracted by performing the inverse
Laplace transformation which boils down to taking $z=i0^{+}$.

Hence, the spin current in the direction $\nu\in\left\{ x,y,z\right\} $
can be obtained by substituting into \eqref{eq:observable-kernel}
the following as observable:
\begin{align*}
I_{\textrm{spin},\nu}^{r} & =\frac{d}{dt}\sum_{k,\sigma\sigma'=\uparrow,\downarrow}\sigma_{\sigma\sigma'}^{\nu}c_{kr\sigma}^{\dagger}c_{kr\sigma'}=-i\sum_{k,\sigma\sigma'=\uparrow,\downarrow}\sigma_{\sigma\sigma'}^{\nu}\left[c_{kr\sigma}^{\dagger}c_{kr\sigma'},H_{\textrm{tun}}\right]_{-}\\
 & =i\sum_{k,\sigma\sigma'\sigma''=\uparrow,\downarrow}d_{\sigma''}^{\dagger}t_{\sigma''\sigma'}^{r}\sigma_{\sigma'\sigma}^{\nu}c_{kr\sigma}+\textrm{h.c.}
\end{align*}
Analog the charge current is obtained by replacing $\sigma_{\sigma'\sigma}^{\nu}\to\delta_{\sigma'\sigma}$.
As neither charge nor spin current add any additional energy dependence
to the observable kernel \eqref{eq:observable-kernel} the same integrals
needs to be solved for these currents as for the computation of the
decay rates via the effective Liouvillian \eqref{eq:full-Liouvillian}
discussed prior.

\begin{figure}[t]
\begin{centering}
\includegraphics[width=0.45\columnwidth]{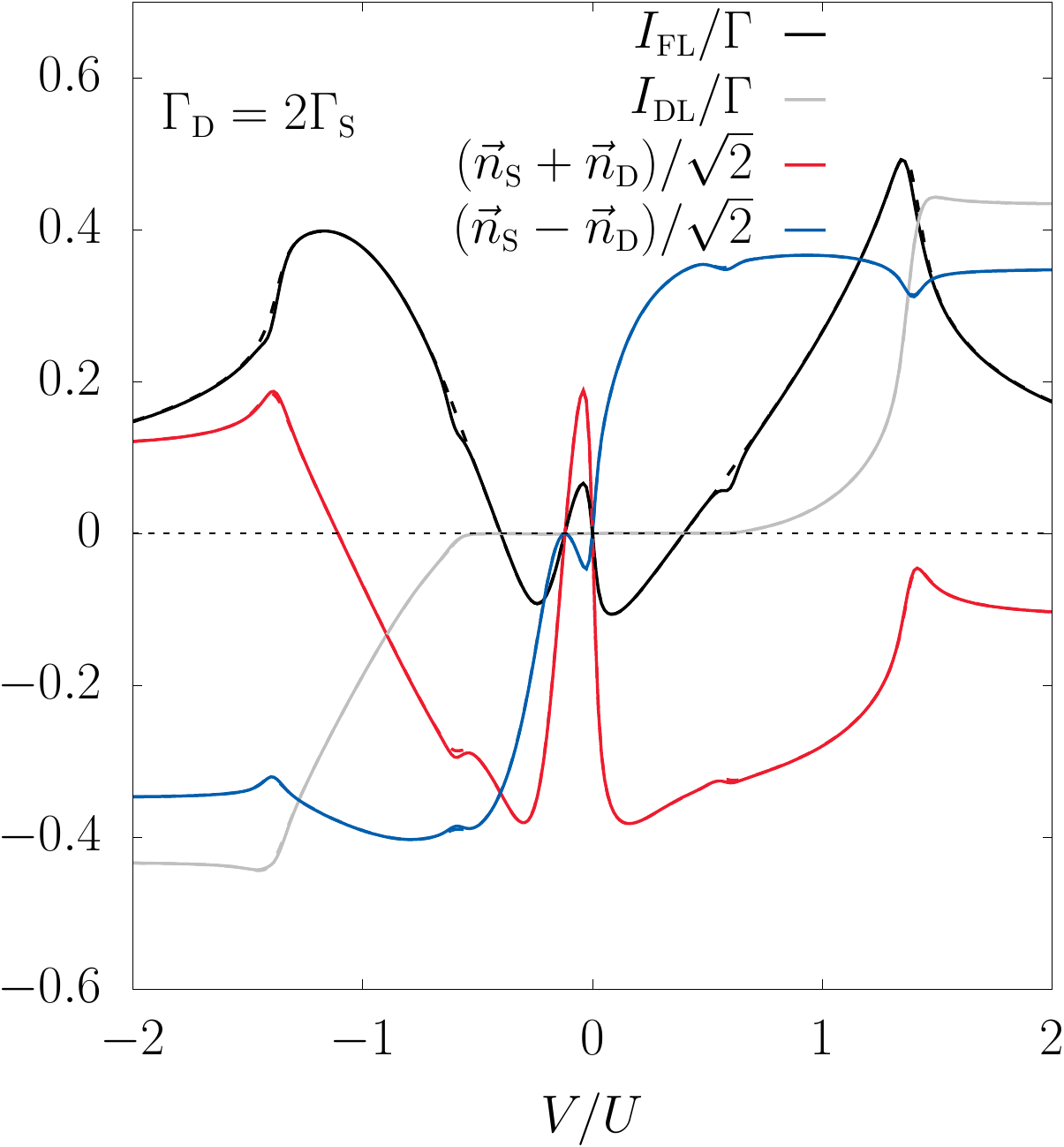}\hspace{0.5cm}\includegraphics[width=0.45\columnwidth]{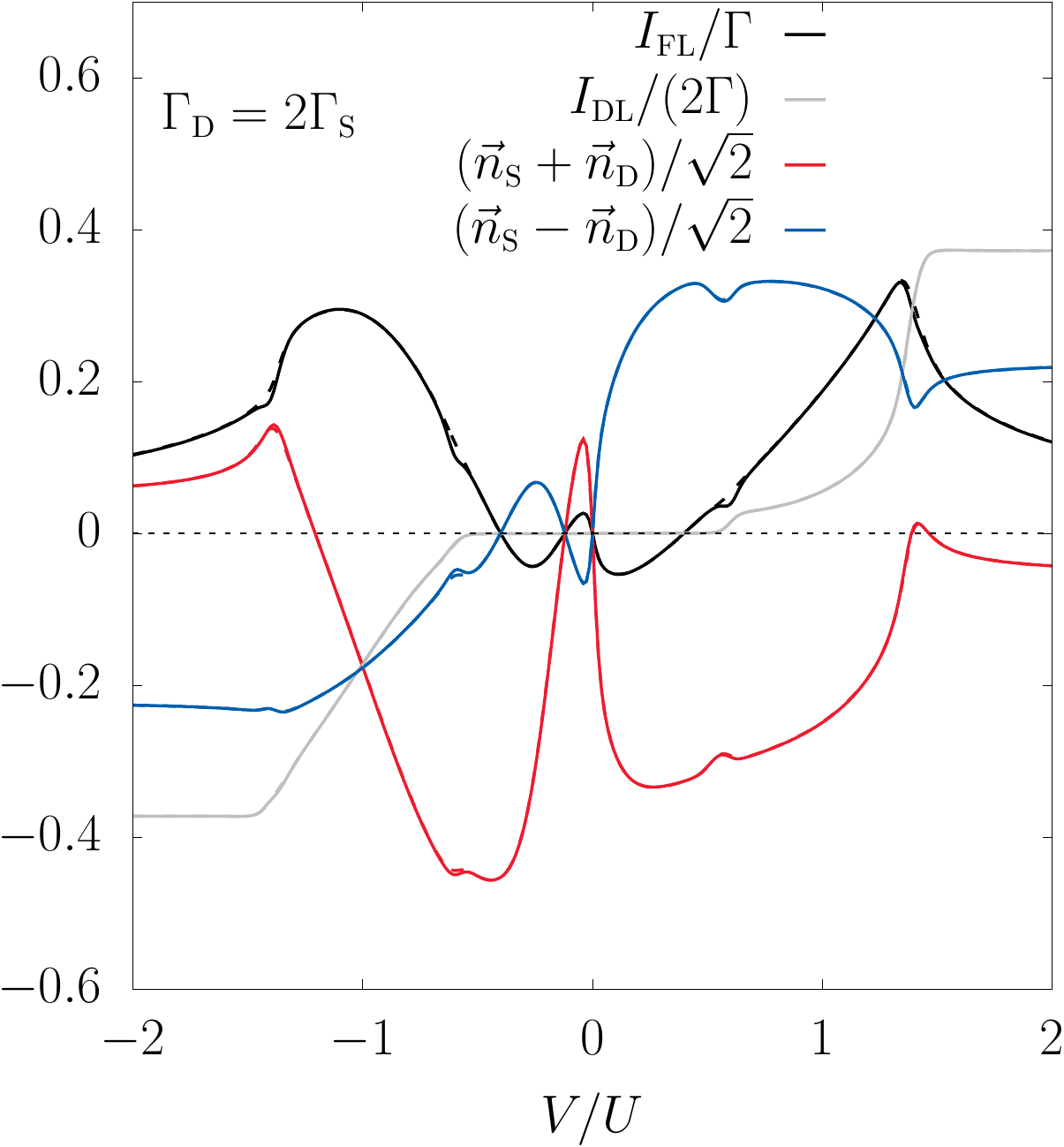}
\par\end{centering}

\caption{Spin torques and the spin projection of the stationary dot occupations in the directions $\vec{n}_{\textrm{S}}\pm\vec{n}_{\textrm{D}}$
for fixed $\tilde{\varepsilon}=0.2U$, other parameters are $U=10^{2}T=10^{3}\Gamma$.
The left plot is for $\theta=\pi/2$ and the right for $\theta=\pi/4$.
The full lines include the induced magnetic field only in $\mathcal{O}\left(\Gamma\right)$,
but still take all $\mathcal{O}(\Gamma^{2})$ decay rates into account.
The dashed lines where computed by including the full $\mathcal{O}\left(\Gamma^{2}\right)$-corrections.
Most of the dashed lines are not visible as they coincide with the
full lines. Overall, no significant, qualitative changes occur; especially
with respect to the sign of the field like torque which is the crucial
ingredient for the effects under consideration.\label{fig:check}}
\end{figure}

\subsection{$\mathcal{O}(\Gamma)$-approximations for the effective magnetic
field}

All plots shown in the main letter use an $\mathcal{O}(\Gamma)$-approximation
of the effective magnetic field., i.e. the full $\mathcal{O}(\Gamma)$-contribution to the Liouvillian 
as given at Eq.~\eqref{eq:Leading order} is taken into account. In contrast, in the imaginary part of the Liouvillian all terms in $\mathcal{O}(\Gamma^{2})$ are considered, since they yield the first non-vanishing contribution to the  decay rates. The used exact integral expression for the $\mathcal{O}(\Gamma^{2})$-decay
rates are given in the supplement of Ref.~[\citenum{Gergs15}]. In Fig.~ \ref{fig:check}  we show the spin torques and stationary dot occupation for two exemplary cuts through the Coulomb diamond. We also show the results when including the $\mathcal{O}(\Gamma^{2})$-corrections to the effective magnetic field and observe that no significant changes occur. 

\section{Further analysis of the spin switching behavior}

\subsection{Weaker reservoir polarizations}

\begin{figure}[t]
\begin{centering}
\includegraphics[width=0.48\columnwidth]{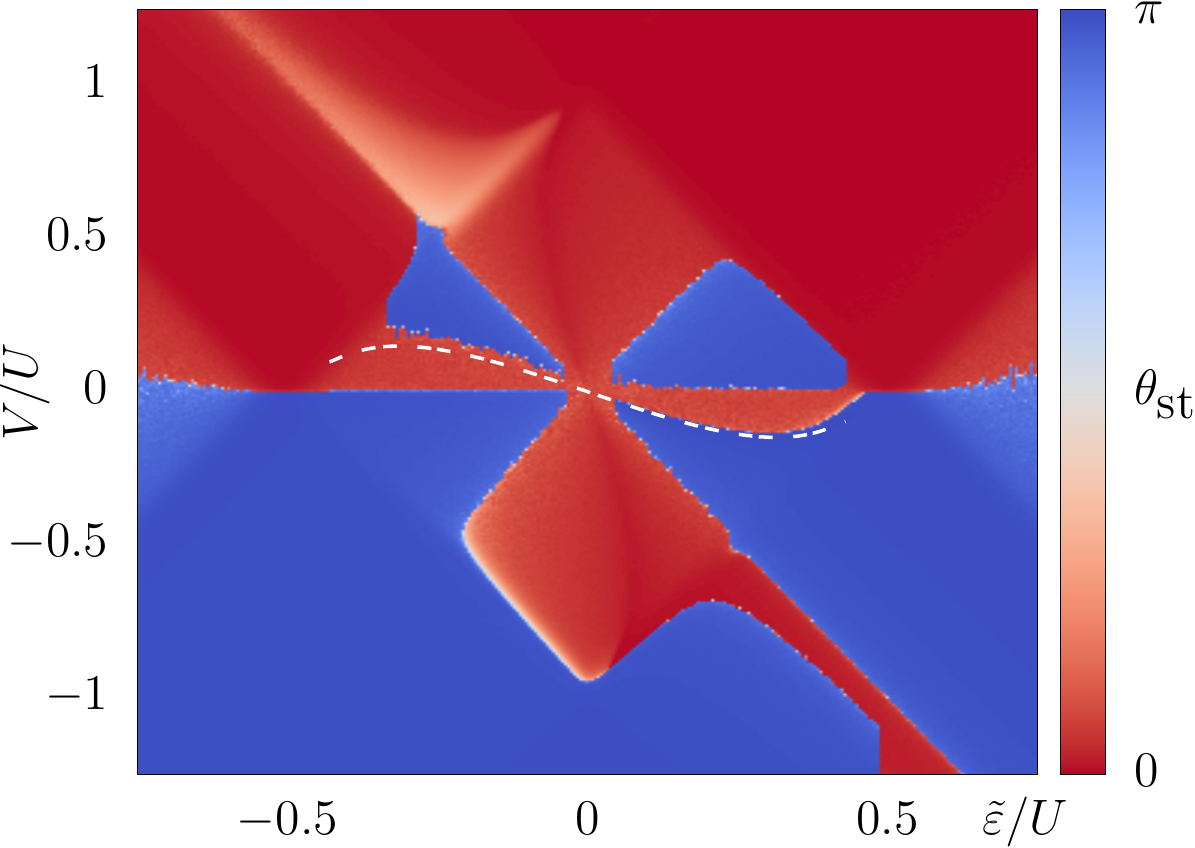}\hspace{0.5cm}\includegraphics[width=0.48\columnwidth]{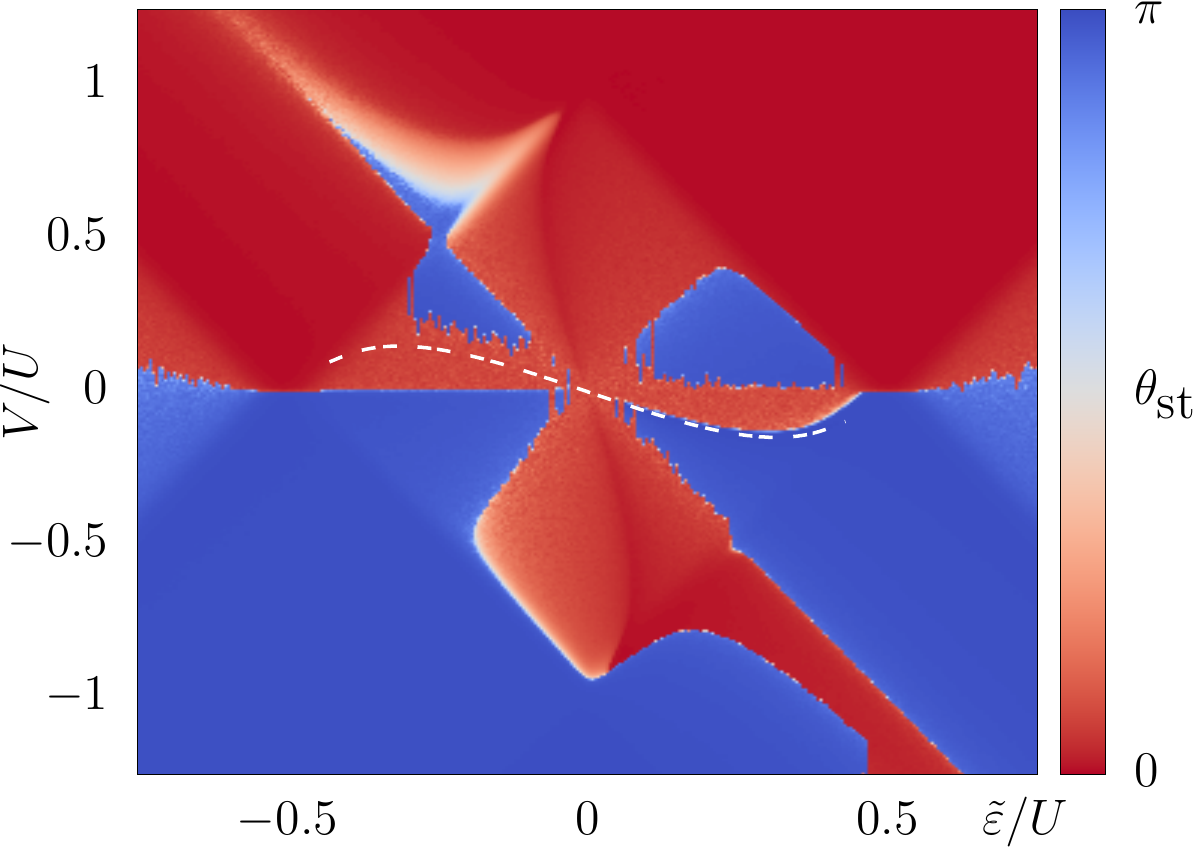}
\par\end{centering}
\caption{Switching behavior for the same parameters as used for Fig. 2 of the
main part, i.e $U=10^{2}T=10^{3}\Gamma$, an easy-axis anisotropy
of $KV=0.01\Gamma$, a Gilbert damping of $\alpha=0.1$ and temperature
fluctuations $\sigma_{\textrm{thermal}}^{2}=10^{-4}$. The left plot
is for reservoir polarizations $p_{r}=0.9$ and the right one for
$p_{r}=0.7$. For $p_{r}=0.7$ some artifacts begin to appear for
the slow transition from parallel to anti-parallel along the resonance
line. Also fluctuations become more pronounced for the parallel configurations,
making the partner resonance $\vec{B}_{\textrm{ind}}\cdot(\vec{n}_{\textrm{D}}+\vec{n}_{\textrm{S}})=0$
close to the $\tilde{\varepsilon}=0$ line visible to the bare eye.\label{fig:polarizations}}
\end{figure}

The main part of this letter considered for clarity only very strongly polarized reservoirs with a polarization of $p=0.99$. However, there
is no fundamental limitation with respect to the polarization strength.
Obviously, considering extremely weak polarizations $p_{r}\to0$ is
not useful as the induced magnetic field \eqref{eq:magnetic-field}
vanishes and therefore also all spin torques. Besides this considering
$p_{r}<0.99$ just yields a reduction of the spin torque strength,
Fig. \ref{fig:polarizations} displays corresponding switching diagrams.
Overall the general properties discussed for this setup are also present
for such lower polarizations. For $p_{r}=0.7$ stronger fluctuations
occur for the parallel configurations and the switching from parallel
to anti-parallel is so slow that significant hysteresis occurs (plots
show scans from $V=-1.25\tilde{\varepsilon}$ to $V=1.25\tilde{\varepsilon}$
with fixed $\tilde{\varepsilon}$). Due to the more pronounced thermal
fluctuations for the parallel configuration the partner resonance
$\vec{B}_{\textrm{ind}}\cdot(\vec{n}_{\textrm{D}}+\vec{n}_{\textrm{S}})=0$
close to the $\tilde{\varepsilon}=0$ line becomes visible as a region
of suppressed thermal fluctuations in the parallel configuration region.
The underlying reason for this supression is a much stronger dampinglike
spin torque (cf. Fig. 4 of the main part) which supresses all fluctuations
away from the stable direction.

\begin{figure}[b]
\begin{centering}
\includegraphics[width=0.5\columnwidth]{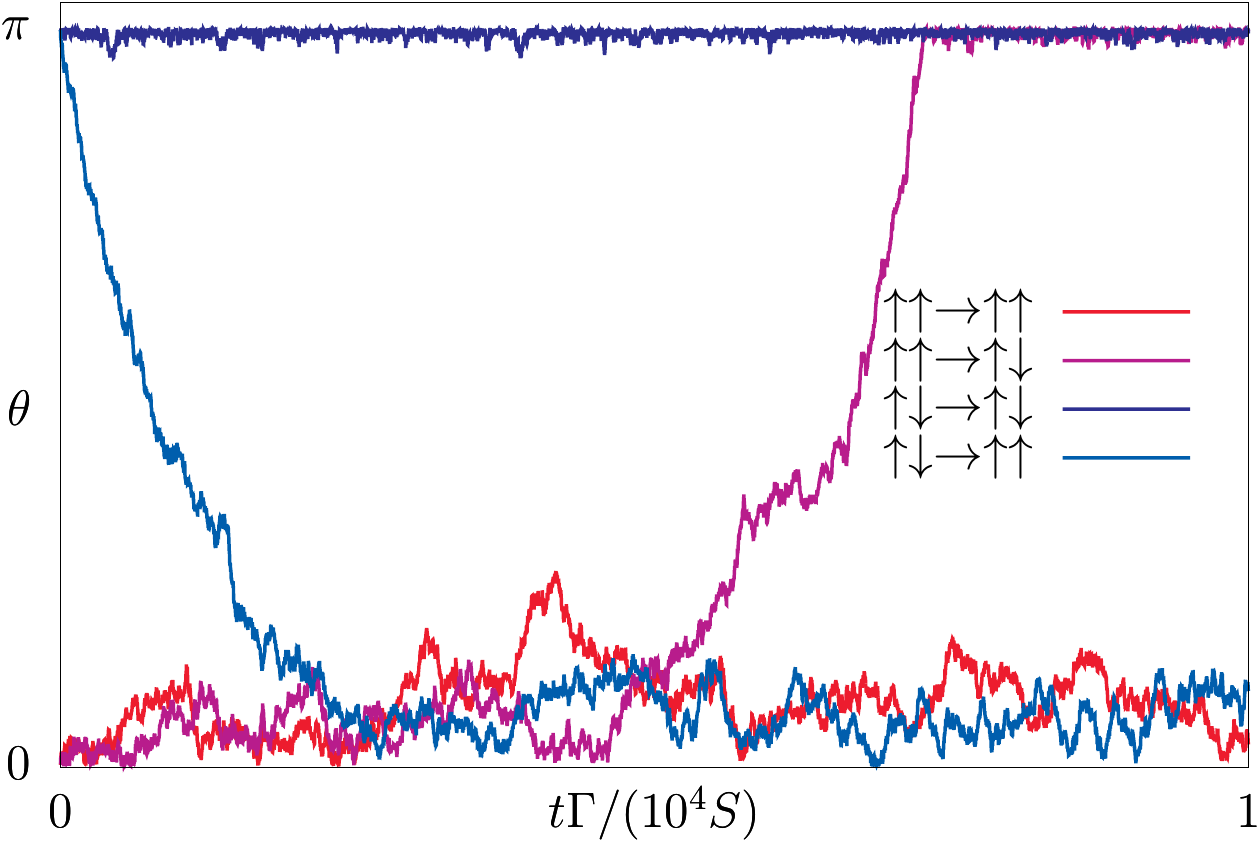}
\par\end{centering}
\caption{Switching time behavior for the same parameters as in Fig. 2 of the
main part of this letter. The switching to parallel corresponds to
the point marked with a filled black dot in that Figure, i.e. $\tilde{\varepsilon}=-0.2U,\,V=0.1U$,
and the switching to an anti-parallel stationary state corresponds
to the hollow dot, i.e. $\tilde{\varepsilon}=-0.2U,\,V=0.15U$. We note that the switching from anti-parallel to parallel (light blue line) is fast, while the opposite switching (violet) is slow since in the latter case thermal fluctuations first have to generate a sufficient opening angle. 
\label{fig:switching}}
\end{figure}

\subsection{Switching behavior}

Fig. \ref{fig:switching} shows a typical switching behavior close
to the resonance. For the case where the initial configuration is
equal to the stationary one only thermal fluctuations occur. They
are strongly suppressed for the anti-parallel configuration
due to the corresponding stronger spin torques (cf. insets of Fig.
4 from the main part) pushing the spin back to an anti-parallel configuration.
For similar reasons the switching from anti-parallel to parallel is
quite fast due to the strong spin torques in the initial anti-parallel
configuration. For the inverse switching process of switching from parallel to anti-parallel one has vanishing
spin torques in the limit of perfect parallel configurations $\theta\to0$. Hence, one observes a rather long waiting time during which thermal fluctuations first have to generate a sufficient initial $\theta_{\textrm{fluctuated}}$
from which on the spin torques take the lead and give a switching
of comparable speed as for the anti-parallel to parallel switching.

\subsection{Experimental estimate for Galfenol}

Let us briefly comment on the effect of choosing a more complicated
magnetic energy instead of the simple easy-axis one. Motivated by
the materials like Galfenol~\citep{ExpValve1,ExpValve5} (iron-gallium
alloys) we have also analyzed the case in which the quantum dot is
attached to two nanomagnets possessing cubic anisotropy 
\begin{equation}
\frac{E_{r}}{V}=-K_{\textrm{cubic}}\left(n_{r,\textrm{x}}^{4}+n_{r,\textrm{y}}^{4}+n_{r,\textrm{z}}^{4}\right)-K_{\textrm{uniaxial}}n_{r,\textrm{z}}^{2}\textrm{ .}\label{eq:cubic}
\end{equation}
As mentioned above, all qualitative effects discussed in this article
still appear. In fact, switching becomes easier compared to an easy-axis
anisotropy; e.g., the required thermal fluctuations to explore the
phase space and initialize the switching process are lower. We obtain
a simple estimate of the appearing energy scales by considering nanomagnets
of $10\textrm{nm}\times10\textrm{nm}\times2\textrm{nm}$ made of $\textrm{Fe}_{78}\textrm{Ga}_{22}$
with parameters~\citep{ExpValve5} $K_{\textrm{cubic}}\approx-8\cdot10^{21}\textrm{ev}/\textrm{m}^{3}$
, $K_{\textrm{uniaxial}}\approx1\cdot10^{21}\textrm{ev}/\textrm{m}^{3}$,
corresponding to the energy scale $K_{\textrm{cubic}}V\approx-1.6\textrm{meV}$.
We studied a setup with tunneling rate $\Gamma=-5K_{\textrm{cubic}}V\approx8\textrm{meV}$,
Coulomb interaction $U=10^{2}T=10^{3}\Gamma$ , thermal variance $\sigma_{\textrm{thermal}}^{2}=10^{-3}$
and Gilbert damping~\citep{ExpValve4} $\alpha=0.017$. Tunneling rates
in that order of magnitude have been realized~\citep{ExpValve6} in
quantum dot setups before; the more challenging aspect is to realize
a sufficiently large Coulomb blockade exceeding this energy regime.
However, even though they are not the standard case, quantum dots
with charging energies up to \ensuremath{\sim} eV have already been
realized~\citep{ExpValve2}.

\begin{figure}[t]
\begin{centering}
\includegraphics[width=0.49\textwidth]{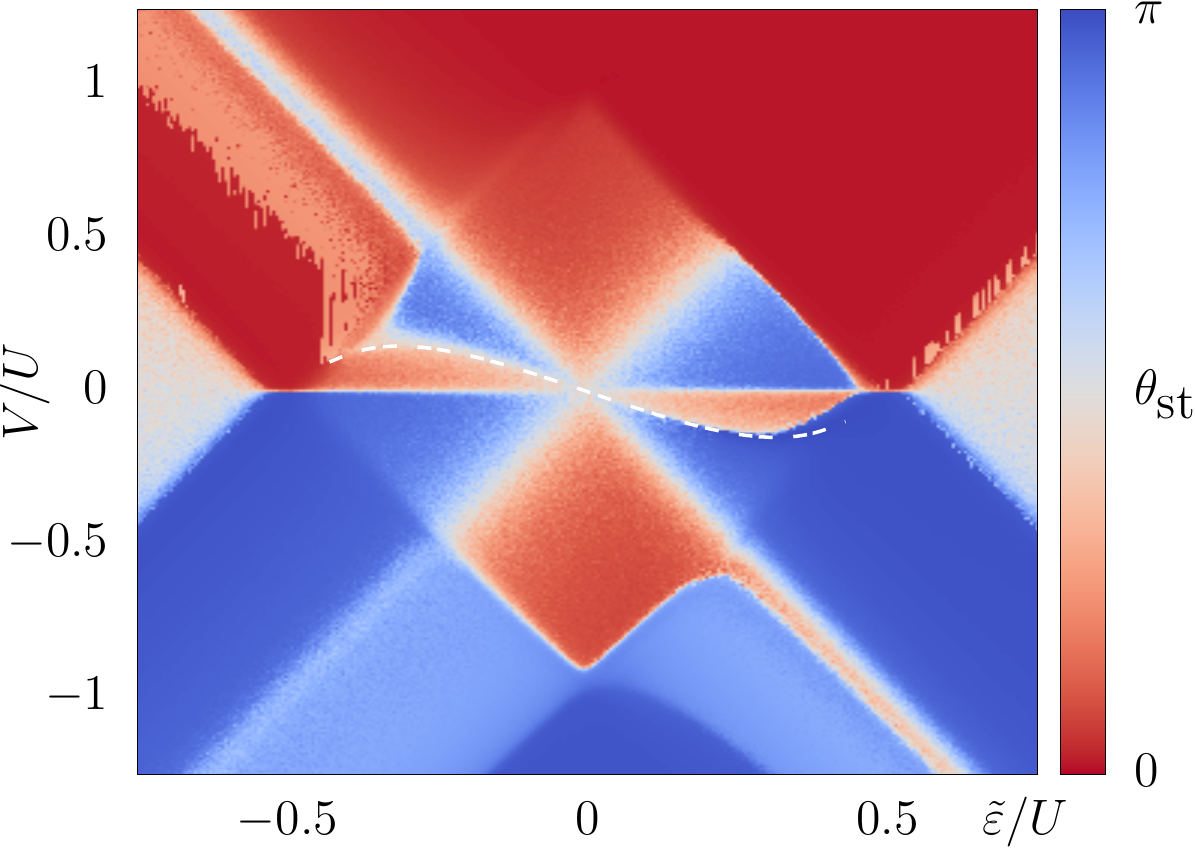}
\par\end{centering}
\caption{Switching diagram displaying the average opening angle $\theta$ after
$t=10^{4}S/\Gamma$ and subsequently averaged over the same time.
Parameters are $U=10^{2}T=10^{3}\Gamma$. Here a cubic anisotropy
\eqref{eq:cubic} with $K_{1}V=0.2\Gamma$, a Gilbert damping of $\alpha=0.017$
and temperature fluctuations $\sigma_{\textrm{thermal}}^{2}=10^{-3}$
have been considered. The universal switching behavior in the Coulomb
diamond remains qualitatively unaltered compared to the simple easy-axis
case considered throughout the main part of this letter.\label{fig:switching-Galfenol}}
\end{figure}

Fig. \eqref{fig:switching-Galfenol} shows the result of this simulations,
all qualitative aspects discussed in the main text are present for this
setup and should be observable. The measurement time to wait for the
switching to happen and to average subsequently over the resulting
configurations whilst measuring used for this plot is given by $t_{\textrm{measure}}=10^{4}S/\Gamma\approx0.8S\cdot\textrm{ns}.$
Hence, for realistic $S\approx10$ for nanomagnets and $S\approx10^{6}$
for magnetic nano pillars, one
obtains switching times of $t_{\textrm{measure}}\approx8\textrm{ns}$
and $t_{\textrm{measure}}\approx800\textrm{\ensuremath{\mu\textrm{s}}}$,
respectively.

 \section{Derivation of Eq.~(5)}  
In this section, we derive Eq.~(5).  First, we generalize Eq.~(1) for $r=D$ to include a small FMR driving field $\vec{h}(t)$ by sending $\vec{H}_D\rightarrow \vec{H}_D+\vec{h}(t)$ on the right-hand side. Supposing the driving field to be of the form $\vec{h}(t)=h_x e^{i \Omega t} \vec{e}_x $, we linearize $\vec{n}_D\approx \delta \vec{n}e^{i \Omega t}+n_D^z \vec{e}_z$, where $n_D^z=\pm 1$ denotes the steady-state fixed point orientation. The response $\delta \vec{n}=n_x \vec{e}_x+n_y \vec{e}_y$, which is generally complex, is obtained by inserting Eq.~(2) into Eq.~(1) and using the linearized form of $\vec{n}_D$, resulting in: 
\begin{equation} \left(\begin{array}{cc} n_D^z\left[\gamma H-n_D^z \alpha Si\Omega-I_{\rm{FL}}^{\rm{D}}\right] & -n_D^z\left[Si\Omega -n_D^z I_{\rm{DL}}^{\rm{D}}\right]\\ -n_D^z\left[Si\Omega-n_D^z I_{\rm{DL}}^{\rm{D}}\right] & -n_D^z\left[\gamma H-n_D^z \alpha Si\Omega-I_{\rm{FL}}^{\rm{D}}\right] \end{array}\right)\left(\begin{array}{c} n_{x}\\ n_{y} \end{array}\right)\equiv \chi^{-1}\left(\begin{array}{c} n_{x}\\ n_{y} \end{array}\right)=\left(\begin{array}{c} h_{x}\\ 0 \end{array}\right)\, . \label{sup1} \end{equation}
Here, $H=B_{\rm{extern}}+KV$ is the effective field, and $I_{\rm{DL}}^{\rm{D}}$ and $I_{\rm{FL}}^{\rm{D}}$ are understood to be evaluated at $\theta=0$ for $n_D^z=+1$ and $\theta=\pi$ for $n_D^z=-1$.  We have neglected the thermal field $\vec{H}_{\rm{thermal}}$, as thermal broadening is usually negligible at low temperatures. Eq.~(\ref{sup1}) can then be inverted to obtain $n_x=n_x(\Omega)=\chi_{xx}(\Omega) h_x$ and hence the time-dependent solution $n_x(\Omega,t)=n_x(\Omega) e^{i\Omega t}$.  Finally, to obtain to full response $n_x(t)$ to a driving field $\vec{h}(t)=h_x \rm{cos}(\Omega t)$, we add the positive and negative frequency solutions: $n_x(t)=(n_x(\Omega,t)+n_x(-\Omega,t))/2=\Re[n_x(\Omega,t)]= \chi'_{xx}h_x \rm{cos}(\Omega t)+\chi''_{xx}\rm{sin}(\Omega t)$, where $\chi'_{xx}=\Re{\chi_{xx}}$ and $\chi''_{xx}=-\Im{\chi_{xx}}$, which is written explicitly in Eq.~(5).

\end{document}